\documentclass[%
 reprint,
 amsmath,amssymb,
 aps,
10pt,prd,nofootinbib,floats,floatfix,amsfonts,preprintnumbers,notitlepage,superscriptaddress,twocolumn]{revtex4-2}

\usepackage{subfigure}
\usepackage{graphicx}
\usepackage{dcolumn}
\usepackage{bm}
\usepackage{comment}
\usepackage{color}

\newcommand{\orcid}[1]{\href{https://orcid.org/#1}{\includegraphics[scale=0.15]{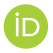}}}
\begin{document}
\preprint{DOI:10.1103/PhysRevD.106.063025}

\title{Figures of merit for a stochastic gravitational-wave background measurement by LISA:\\Implications of LISA Pathfinder noise correlations}
\author{Guillaume Boileau \orcid{0000-0002-3576-69689}}
\affiliation{Universiteit Antwerpen, Prinsstraat 13, 2000 Antwerpen, Belgium}
\affiliation{%
Universit\'{e} C\^{o}te d'Azur, Observatoire de la C\^{o}te d'Azur, CNRS, ARTEMIS, CS 34229, F-06304 Nice Cedex 4, France
}%
 \email{guillaume.boileau@oca.eu}
\author{Nelson Christensen \orcid{0000-0002-6870-4202}}%
 \email{nelson.christensen@oca.eu}
\affiliation{%
Artemis, Observatoire de la C\^{o}te d'Azur, Universit\'{e} C\^{o}te d'Azur, CNRS, CS 34229, F-06304 Nice Cedex 4, France
}%
\author{Renate Meyer \orcid{0000-0003-0268-8569 }}
\email{renate.meyer@auckland.ac.nz}
 \affiliation{
 Department of Statistics, University of Auckland, 1010 Auckland, New Zealand}

\date{\today}

\begin{abstract}
An important goal of the Laser Interferometer Space Antenna (LISA) is to observe a stochastic gravitational-wave background (SGWB). A study of possible correlated noise in LISA is relevant to establish limits for this future measurement. To test noise investigation methods under somewhat realistic conditions, we use the data of LISA Pathfinder. We calculate the coherence between the LISA Pathfinder differential acceleration of the two test masses with magnetic fields, temperature, and micronewton cold gas thruster activity in the spacecraft. We apply our observed correlations to LISA, and estimate how the presence of such correlated noise would affect its search for a SGWB. In the context of a figure of merit, we estimate the effect of noise on the LISA SGWB search. 

\end{abstract}

\maketitle

\section{Introduction}

The Laser Interferometer Space Antenna (LISA) will attempt to detect the stochastic gravitational wave background (SGWB)~\cite{Christensen_2018} in the frequency band $[1.10^{-5}$, $1]$ Hz (or possibly more restrictively to $[1.0^{-4}$, $0.1]$ Hz). The SGWB will be present in the two LISA time delay interferometry (TDI) science channels, $A$ and $E$~\cite{PhysRevD.100.104055,Tinto2005}. A basic assumption is that the noise in the LISA TDI $A$, $E$, and $T$ channels are uncorrelated.
We want to study the ramification of correlated noise entering into the LISA data and the effects it would have on the detection of a SGWB, plus the estimation of its parameters.

The presence of nonideal LISA noise will presumably affect the ability to conduct spectral separation of the SGWB. We have demonstrated such spectral separation in previous studies~\cite{PhysRevD.103.103529,2021arXiv210504283B,Boileau:2021gbr}. For this present study, we examine some sources of noise on the spacecraft that can potentially disturb the LISA measurement of the SGWB. 
We conduct these noise studies with real data, namely from LISA Pathfinder (LPF)~\cite{LPFscience, PhysRevLett.116.231101}. We search for correlations between the difference of acceleration of the two LPF test masses $\Delta g $, and three other physical quantities: the magnetic field in the satellite, the temperature, and the micronewton cold gas thruster ($\mu$ thruster) activity~\cite{PhysRevD.99.122003, PhysRevD.98.102005}. The $\mu$ thrusters are the satellite's orbit-maintaining system so that the satellite preserves and protects a test mass in free fall. Admittedly, the LPF satellite and optical system are different than those proposed for LISA. However, the LPF data does provide an important opportunity to explore noise sources that could be similar to those of LISA. Our use of LPF data to investigate possible issues with the LISA performance is similar to a recent study that investigated LPF noise transients (glitches) and how similar events in the LISA data could influence gravitational-wave searches~\cite{PhysRevD.105.042002}.

LPF was located at Lagrange point L1, about 1.5 million kilometers from Earth in the direction of the sun. It is a stable position, an object at this point follows the orbit of the Earth without adding orbital energy. But, the position of the spacecraft on point L1 must be addressed. Space engines must maintain a constant flow of solar energy (ensure proper orientation of solar panels) and it is important to ensure system stability. To ensure this, each spacecraft uses $\mu$ thrusters technology~\cite{PhysRevD.99.122003}, but this produces noise correlated with the difference in acceleration from the test masses. 

We use the public LPF data available on the site \href{http://lpf.esac.esa.int/lpfsa/#results}{LISA Pathfinder Legacy Archive}~\footnote{http://lpf.esac.esa.int/lpfsa/}. This database contains the time series of the difference in acceleration of the mass tests $ \Delta g(t)$, but also of other channels such as the temperature of the instrument, magnetic field measurements, or the strengths of the action of the $\mu$ thrusters. There is a catalog of different levels of data processing of $ \Delta g (t) $. The levels are named Level 0 to Level 3. 
We calculate the coherence between the measurement of the magnetic field, the temperature and the $\mu$ thrusters in the spacecraft and the differential acceleration of the two test masses ($ \Delta g $). 

We discuss the addition of correlated noise in the context of LISA, and we try to detect a simulated SGWB.
Such a detection will be the result of the observation of the SGWB in the two science channels, $ A $ and $ E $. We also use the $ T $ channel as the control channel.
With the addition of correlated noise we develop new figures of merit.
We will also test the effect of a localized increase in noise from a LISA MOSA (movable optical subassembly). 

The organization of this paper is as follows. In Sec.~\ref{sc:LPF} we describe the observed LPF correlations that will be used as a worst-case scenario for LISA. In Section~\ref{sc:LPF+LISA}, we present the injection on LPF correlations into the LISA noise budget. The study using spectral separation as a means to observe a cosmological SGWB with LISA is then applied, and given in Sec.~\ref{sc:spectralseparatin}. In Sec.~\ref{sc:ModLISA} we discuss the effect of the injection of a specific correlated noise on a articular LISA satellite, and observe the effects on the SGWB measurement as a "figure of merit". A conclusion is given in Sec.~\ref{sec:conclusion}.

\section{LPF Correlation study: worst case}\label{sc:LPF}

To study the correlation, we introduce the coherence function (also known as the magnitude-squared coherence) $ C_{xy}(f) $, defined below~\cite{welch1967use}. We work in the frequency domain.  The coherence function is a real-valued function of frequency with an absolute value between 0 and 1; we will subsequently refer to it as the coherence. For two time series, $(x(t), y(t))$ the coherence at frequency $f$ is defined by
  \begin{equation}
        C_{xy}(f) = \frac{\left| P_{xy}(f) \right|^2}{P_{xx}(f) P_{yy}(f)} ~ ,
    \end{equation}
   where $ P_{xy}(f) $ denotes the complex-valued cross power spectral density and $ P_{xx}(f)$and $P_{yy} (f) $ the power spectral density (PSD) of the two time series $x(t)$ and $ y(t)$, respectively. It can be thought of as the frequency-domain version of the squared cross-correlation. The coherence at frequency $f$  measures the contribution to the squared correlation coefficient, in analogy to the interpretation of the power spectral density at $f$ as the contribution to the overall variance.

\subsection{Differential acceleration of the LPF test masses in free fall}

For LPF, the first test mass is in free fall, with the satellite movement adjusted by the $\mu$ thrusters to track the mass motion. The second test mass is in pseudo free fall, with its position adjusted by actuation. A laser interferometric system measures the relative acceleration between the two test masses. 
LPF cannot be a gravitational wave detector, simply with its test mass system~\cite{PhysRevD.97.122002, PhysRevLett.123.111101}. But, it can be integrated into a long arm detector like LISA. It is an important opportunity to study LPF and its parasitic noises to understand possible sources of noise in the future LISA mission. 
The two LPF test masses are quasicubic, with a size of $ 46 \pm 5 \times 10^{-3} $ mm and a mass of $ 1.928 \pm 1 \times 10^{-3} $ kg. The nominal distance is $ 376 \pm 5 \times 10^{-2} $ mm.
The differential acceleration of the test masses from the interferometric measurements is affected by several effects: command forces on the test masses, the centrifugal force, spacecraft motion along other degrees of freedom, asymmetric drag free control, Euler force, etc. 
The relative acceleration data is given according to different levels of cleaning, from level 0 (least cleaned) to level 3 (most cleaned); see the
 the LPF Science Archives~\href{http://lpf.esac.esa.int/lpfsa/#results}{website} for access to LPF data) ~\cite{Hewitson_2009, enlighten115284}. For our study we use level 0 in order to capture most noise effects, although we use level 3 for noise transient (glitch) studies.
 We use the data from February13–March 1, 2017. This is a noise measurement phase, in other words, there is no change or manipulation of the parameters of the spacecraft. The dataset can be considered as stationary. The differential acceleration data is a file with a duration of 17 days, 13 hours, 59 minutes, and 59.40 s with a sample rate of 10 Hz.

\subsection{Magnetic field of LPF}

\subsubsection{Magnetic field data from LPF}
We use 12 LPF magnetometers (Table~\ref{tab:mag_sensor}); the sensors are distributed at four locations. In all locations, there are three magnetometers, corresponding to the measurement of the magnetic field along the three spatial dimensions $ x, y, z $. To reconstruct the norm of the magnetic field $ \Vec{B} $ in $ \mu T $ we calculate
\begin{equation}\label{eq:B}
\left|\Vec{B}_i\right|  = \sqrt{B_{i,x}^2+B_{i,y}^2+B_{i,z}^2} ~ .
\end{equation}
There are two types of magnetometers: the "$P$" correspond to magnetometers placed above the test masses, and the "$M$" below, following the direction $ z $ of the satellite ~\cite{Armano_2020}.
\begin{table}[htbp]
\centering
\begin{tabular}{|c|c|c|c|c|}
\hline
& $M_X$ & $M_Y$ & $P_X$ & $P_Y$ \\ \hline
$B_x$ & LDT10286 & LDT10276 & LDT10279 & LDT10283 \\ \hline
$B_y$ & LDT10287 & LDT10277 & LDT10280 & LDT10284 \\ \hline
$B_z$ & LDT10288 & LDT10278 & LDT10281 & LDT10285 \\ \hline
\end{tabular}
\caption{Distributions and names of the magnetic sensors in the LPF satellite~\cite{DiazAguilo:2009cs}}
\label{tab:mag_sensor}
\end{table}
We reconstruct the 16-day time series 
with a sample rate of $ 0.2083 $ Hz. Necessarily, we have to change the sampling frequency of the magnetic field measurements to match the acceleration. 

\subsubsection{Power spectral density of the magnetic field}

\begin{figure}[t] 
    \centering
    \includegraphics[width=\linewidth]{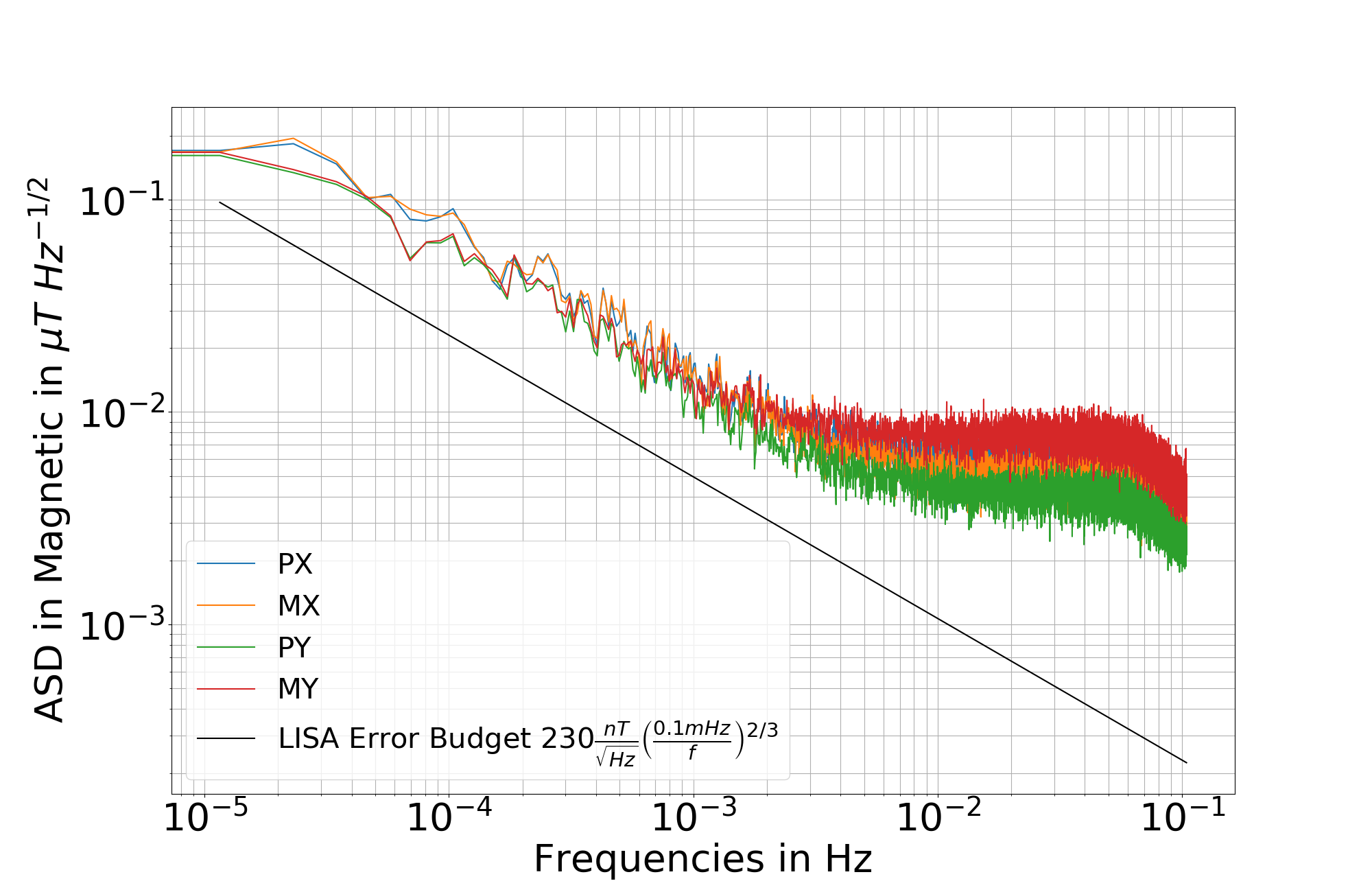}
    \caption{Square root of the average of 17 days of the ASD of all magnetic field sensors in LPF spacecraft, giving the ASD. The dark line is the LISA requirement from the LISA error noise budget \cite{LISAperfo}. The magnetic field should be less than $230 \frac{\rm nT}{\sqrt{Hz}} \left(\frac{0.1 {\rm mHz}}{f} \right)^{2/3}$.  }
    \label{fig:stack_ASD}
\end{figure}

We separate the data into one-day segments. We average the periodograms of the magnetic sensors distributed across the spacecraft to reduce fluctuations. We use the same process for the differential acceleration. The amplitude spectral density (ASD) of the magnetic field for each of the three channels of a fluxgate \cite{Ca_izares_2009} magnetometer has been studied before the LPF mission. In Fig.~\ref{fig:stack_ASD} we show the ASD of the magnetic field during the period of 17 days of all four channels of the Table~\ref{tab:mag_sensor}. In comparison with Fig. 3 of Armano \textit{et. al}~\cite{Armano:2019ekt}, the two plots are similar. The dark line is the LISA requirement. According to \cite{10.1093/mnras/staa830}, the low frequency magnetic field is from the interplanetary magnetic field; the origin of the higher frequency magnetic field fluctuations corresponds to the environment of the spacecraft. There is also an effect of the amplitude of the magnetic field noise with the speed of solar flare. The speed of the solar wind increases the amplitude of this noise. The high frequency of the magnetic field is generated by internal effects. 
We also have an ASD estimate, which is below the requirement for LISA. This is provided by the LISA error budget~\cite{LISAperfo}, which is $ 230 \frac{\text {nT}}{\sqrt{\text {\rm Hz}}} \left (\frac{0.1 \text{mHz}}{f} \right)^{2/3} $ for the LISA test mass environment.    

\subsubsection{Coherence between the magnetic field and the acceleration}

\begin{figure}[htbp]
\centering
\includegraphics[width=\linewidth]{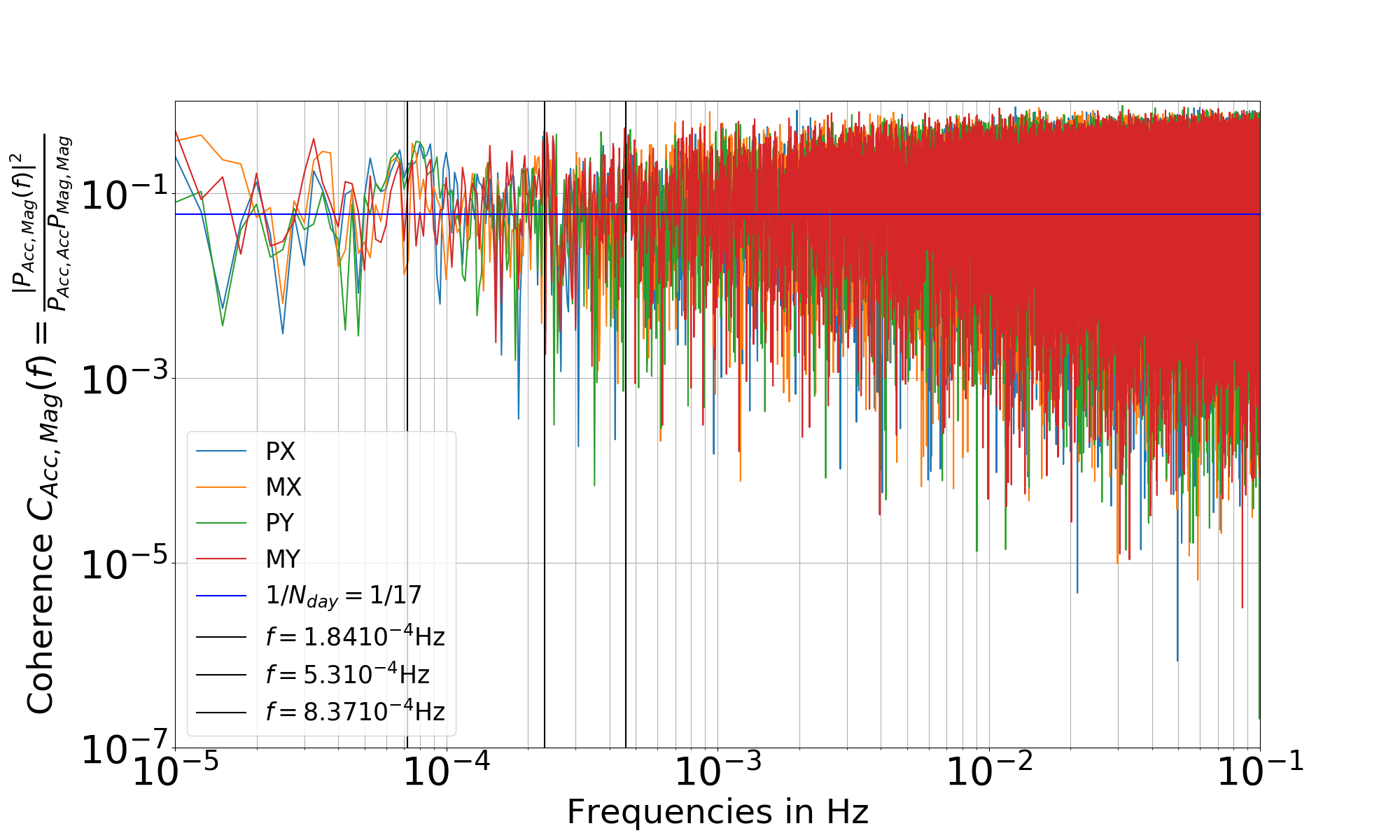}
\caption{17-day average of the coherence of all magnetic sensors in the spacecraft and the acceleration difference of the LPF test masses. The vertical lines correspond to peaks for the coherence. An enlargement of the peaks is given in Fig.~\ref{fig:coherence_mag_zoom}}
\label{fig:coherence_all}
\end{figure}

Here we present the calculation of the coherence between the magnetic field and the differential acceleration of the LPF test masses; this is displayed in Fig.~\ref{fig:coherence_all}. We divide the total time interval into $N=17$ segments, each corresponding to a duration of 1 day. The coherence is flat around the value $ \frac{1}{17}$ (as expected for the number of averages, blue horizontal line). In the case of Gaussian noise, it can be shown that the uncorrelated part of the coherence fluctuates around the value of the inverse of the number of segments used for averaging.
In Fig.~\ref{fig:coherence_all} one can distinguish three peaks in the coherence 
at frequencies $ f = 7.2 \times 10^{-5} $ Hz, $ f = 2.3 \times 10^{-4} $ Hz, and $ f = 4.6 \times 10^{-4} $ Hz (black vertical lines).
We do not have an explanation for what caused these peaks, but only assume that there is some magnetic noise at these levels and study the consequences if such noise were present in LISA.

\begin{figure}[htbp]
\centering
\includegraphics[width=\linewidth]{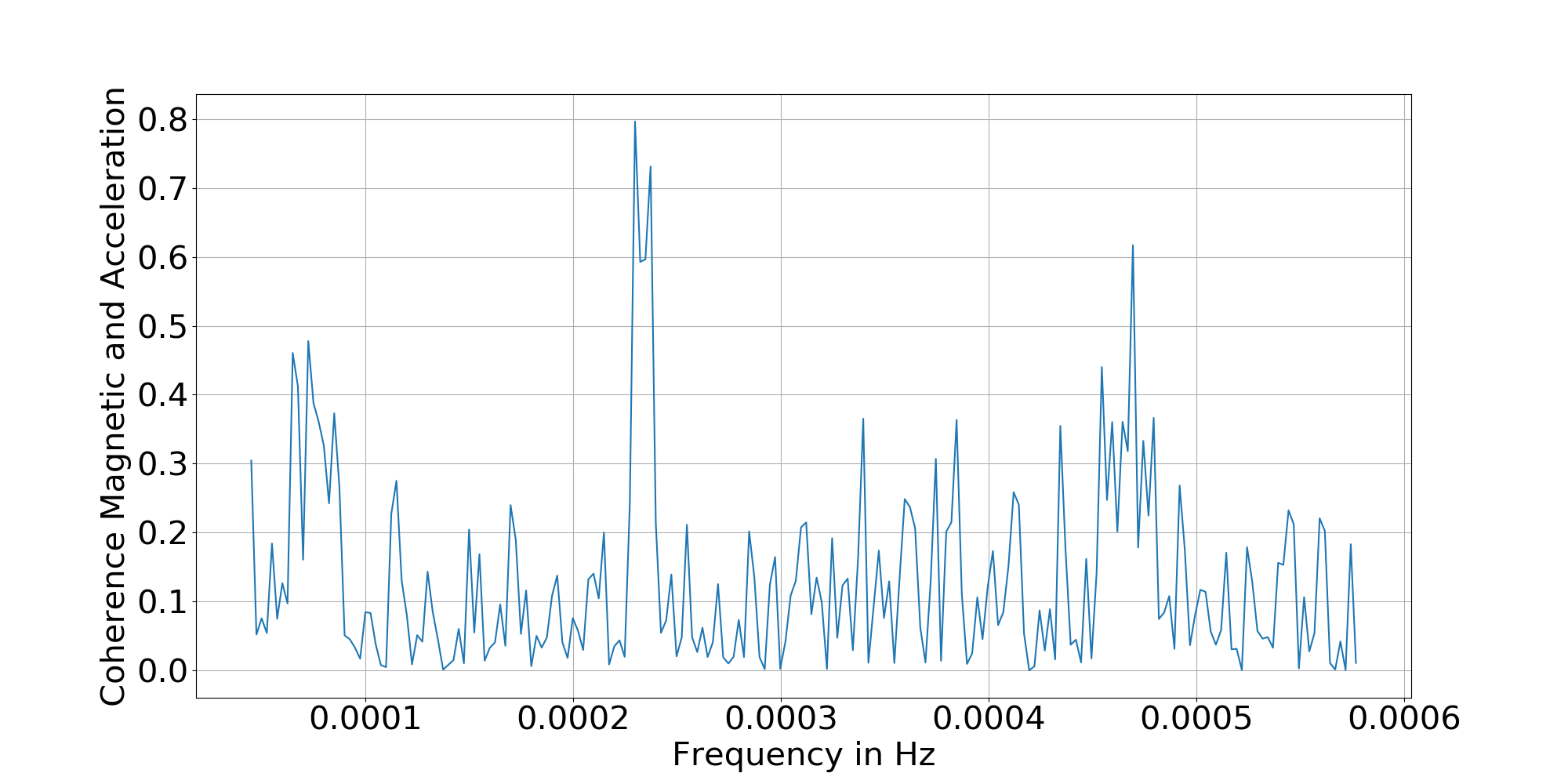}
\caption{Enlargement of the coherence after averaging of all the sensors for the magnetic field and the difference in acceleration throughout  17 days. The three peaks are at the locations of the three vertical lines in Fig.~\ref{fig:coherence_all}. }
\label{fig:coherence_mag_zoom}
\end{figure}

\subsection{$\mu$ thrusters }

In LISA Pathfinder, the Space Technology 7 Disturbance Reduction System~\cite{PhysRevD.98.102005} is used to ensure the stability of the spacecraft. It is the $\mu$ thruster system based on micro-Newtonian cold gas propulsion. The system monitors the position of the satellite in continuous operation throughout the experiment, so the generated noise is present on the difference in acceleration of the two test masses throughout the mission. The sources of disturbances are multiple, such as the $\mu$ thrusters system shot noise, the $\mu$ thrusters system flutter noise, the solar force noise, the radiometer, the magnetic field noise, micrometeoroid impacts or the test mass force noise \cite{PhysRevD.98.102005}. The $\mu$ thrusters could also change the LISA armlengths and thereby affect this analysis; for this study we are ignoring this possibility. 

\subsubsection{LPF $\mu$ thrusters data }
We use the data from 12 $\mu$ thrusters, separated into two groups of six $\mu$ thrusters. The first six $\mu$ thrusters control the six degrees of freedom (three translations and three rotations) during the first part of the mission (until the cold gases contained in their respective tanks are exhausted). Then, during the second part of the mission, the other six $\mu$ thrusters are used. The size of the gas tank is the main limit to the mission time duration. 

For the study of the correlation between the $\mu$ thruster sensors and the differential acceleration we use the $\mu$ thrusters summarized in Table~\ref{tab:thruster_sensor}. The $\mu$ thrusters are distributed in three localities. 

We also study the data from February 13-March2, 2017; for the $\mu$ thrusters we use six thruster sensors (94 to 99 in the LPF data of thrusters of the cold gas system). 
We reconstruct the 17-day time series. The sample rate is $ 1 $ Hz. To study the target channel and the witness channel, we interpolate the thruster channels to change the sample rate, but we cannot study frequencies above the sample rate of $\mu$ thrusters. 

\begin{table}[htbp]
\centering
\begin{tabular}{|c|c|}
\hline
$\mu$ thruster sensors & Position sensors  \\\hline
\hline
GST10094 & Thruster 1 \\ \hline
GST10095 & Thruster 2 \\ \hline
GST10096 & Thruster 3 \\ \hline
GST10097 & Thruster 4 \\ \hline
GST10098 & Thruster 5 \\ \hline
GST10099 & Thruster 6 \\ \hline
\end{tabular}
\vspace{1cm}
\caption{The $\mu$ thrusters in LPF used in this study \cite{Armano:2019ekt}.}
\label{tab:thruster_sensor}
\end{table}

\begin{figure}[htbp]
\centering
\includegraphics[width=\linewidth]{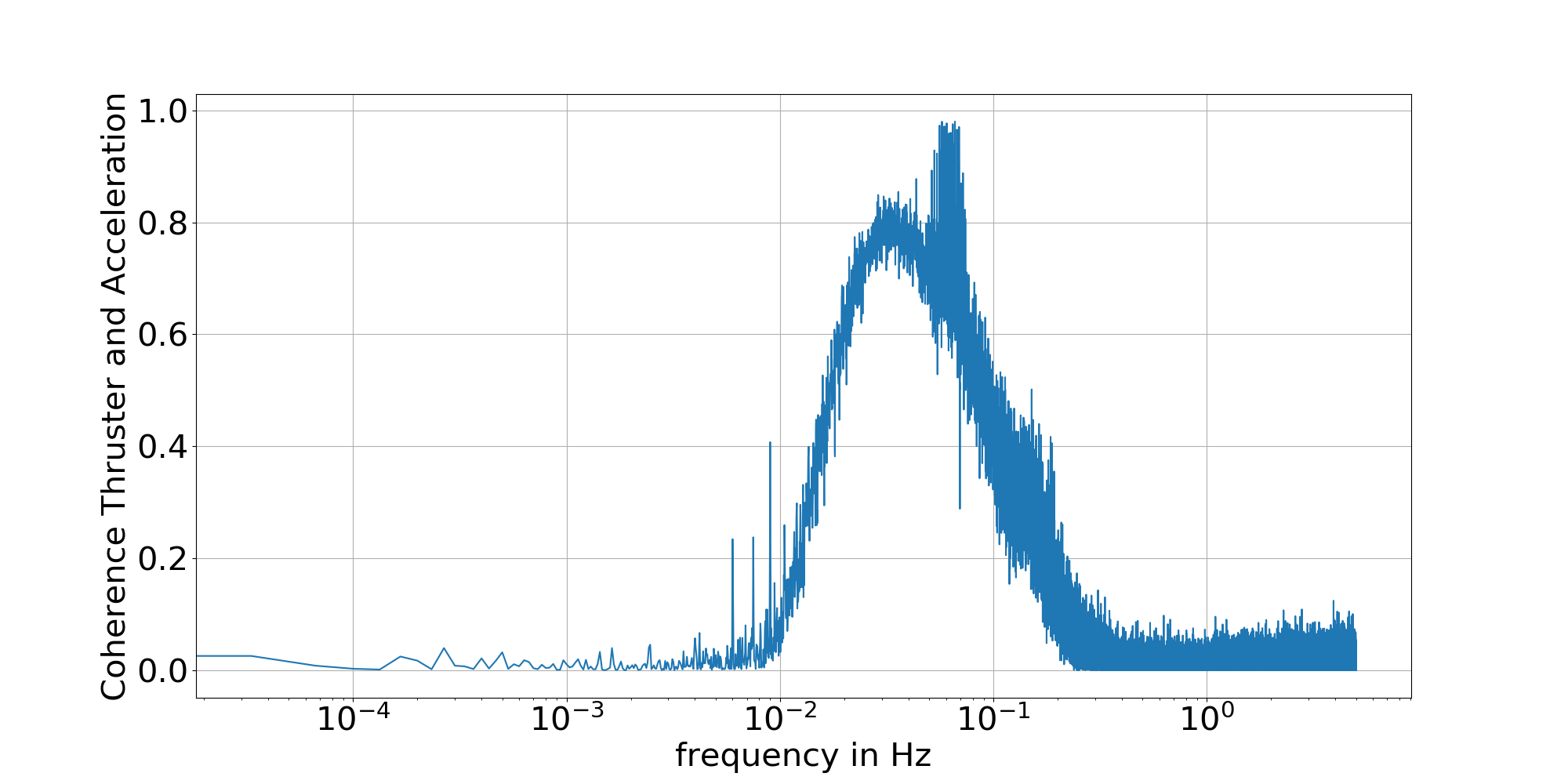}
\caption{Average of the coherence between the $\mu$ thrusters and the differential acceleration of the test masses for 17 days.}
\label{fig:coherence_th}
\end{figure}

\begin{figure}[htbp]
\centering
\includegraphics[width=\linewidth]{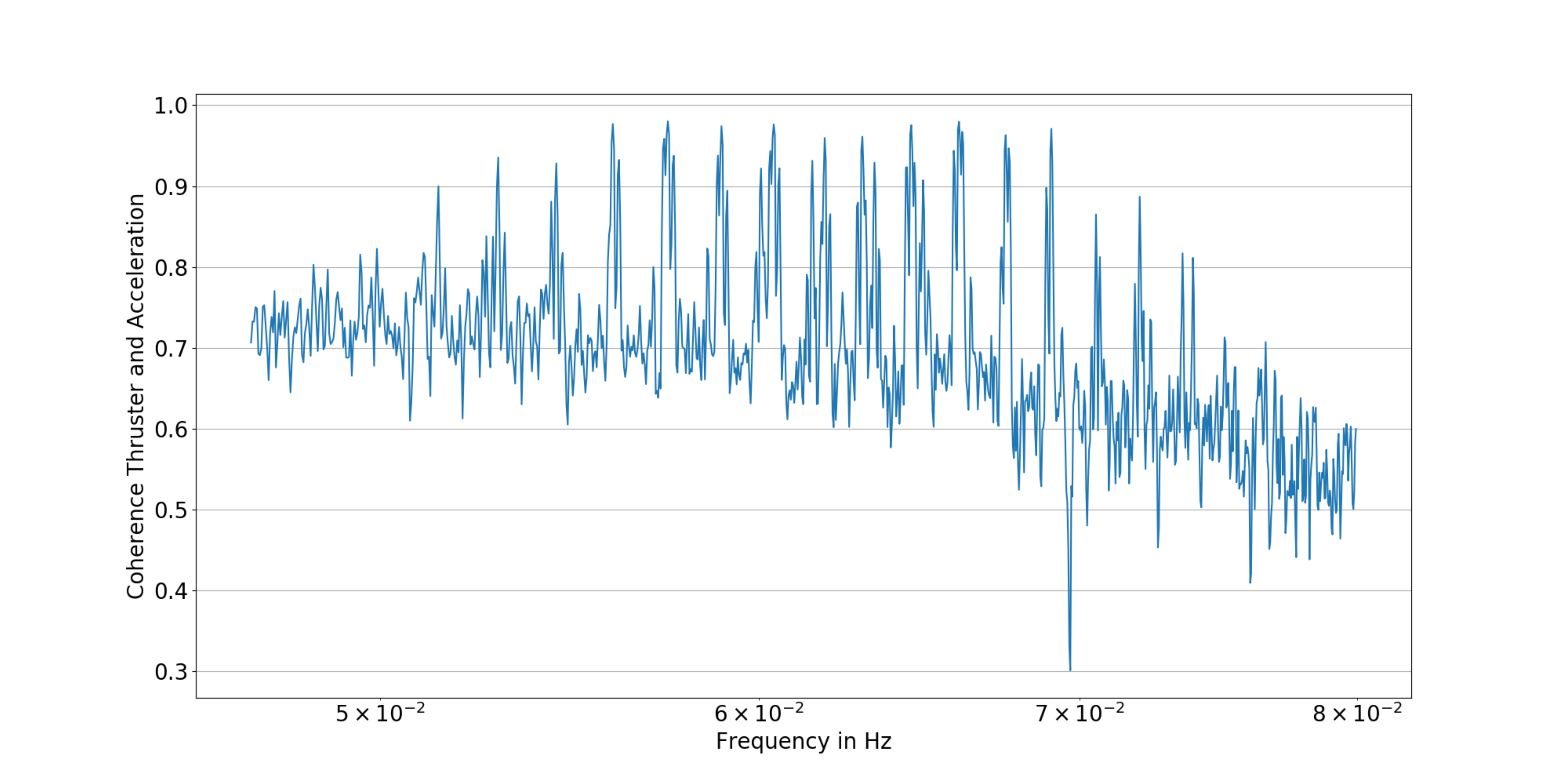}
\caption{Enlargement of the average coherence between the $\mu$ thrusters and the differential acceleration of the test masses for 17 days.}
\label{fig:coherence_th_zoom}
\end{figure}

\subsubsection{Coherence between $\mu$ thrusters and the acceleration}

Figures~\ref{fig:coherence_th} and \ref{fig:coherence_th_zoom} display the coherence over the whole frequency domain and the enlargement of the maximum coherence zone, respectively.
We note the presence of a strong coherence in the band $ [0.047 \ \text{Hz}, 0.0793 \ \text{Hz}] $, indeed, this is characterized by a series of peaks spaced by approximately $ 1.5 $ mHz. These peaks are present on the PSD of the differential acceleration, and also on the PSD of the $\mu$ thrusters \cite{PhysRevLett.116.231101,PhysRevD.99.122003,PhysRevD.97.122002}. The six $\mu$ thruster channels study contains the same peak at the same position. We consider the average over the six channels to maximize the measurement. Experiments have been carried out to demonstrate the effects of $\mu$ thrusters on the acceleration channels during an injection. The presence of an effect of temperature variations on the $\mu$ thrusters was also observed~\cite{PhysRevD.98.102005}.  
The effect of the variation of the temperature changes the distance between the peaks~\cite{PhysRevD.99.122003}. The thermal stability should therefore be a crucial point for the LISA mission. 
The Pearson coefficient for the frequency band $[0.047 \ \text{Hz}, 0.0793 \ \text{Hz}] $ is $ 0.98 $. 

\subsection{LPF spacecraft temperature }
We use 24 thermal sensors (Table~\ref{tab:thermal_sensor}), the sensors are divided into six localizations: four temperature sensors in the electrode housing (EH1) of test mass 1 (TM1) [respectively, four in the electrode housing (EH2) of the test mass 2 (TM2)], three in the optical window 1 (OW1) [respectively, three in the optical window 2 (OW2)], four in the optical bench (OB) and six on the structure of the satellite (STR). These are Betatherm G10K4D372 thermistors. 

\begin{table}[htbp]
    \centering
    \begin{tabular}{|c|c|}
    \hline
         Thermal sensors & Positions \\\hline
         \hline
        TS1    &  EH1 \\       \hline
        TS2    &  EH1 \\       \hline
        TS3    &  EH1 \\       \hline
        TS4    &  EH1 \\       \hline
        TS5    &  EH2 \\       \hline
        TS6    &  EH2 \\       \hline
        TS7    &  EH2 \\       \hline
        TS8    &  EH2 \\       \hline
        TS9    &  OW1 \\       \hline
        TS10   &  OW2 \\       \hline
        TS11   &  OW1 \\       \hline
        TS12   &  EH2 \\       \hline
        TS13   &  OB \\       \hline
        TS14   &  OB \\       \hline
        TS15   &  OB \\       \hline
        TS16   &  OB \\       \hline
        TS17   &  STR \\       \hline
        TS18   &  STR \\       \hline
        TS19   &  STR \\       \hline
        TS20   &  STR \\       \hline
        TS21   &  STR \\       \hline
        TS22   &  STR \\       \hline
        TS23   &  OW2 \\       \hline
        TS24   &  OW1 \\       \hline
 
    \end{tabular}
    \caption{Thermal sensors in LPF used in this study~\cite{Armano:2019ekt}}
    \label{tab:thermal_sensor}
\end{table}

\subsubsection{Power spectral density of temperature}
\begin{figure}[htbp]
    \centering
    \includegraphics[width=\linewidth]{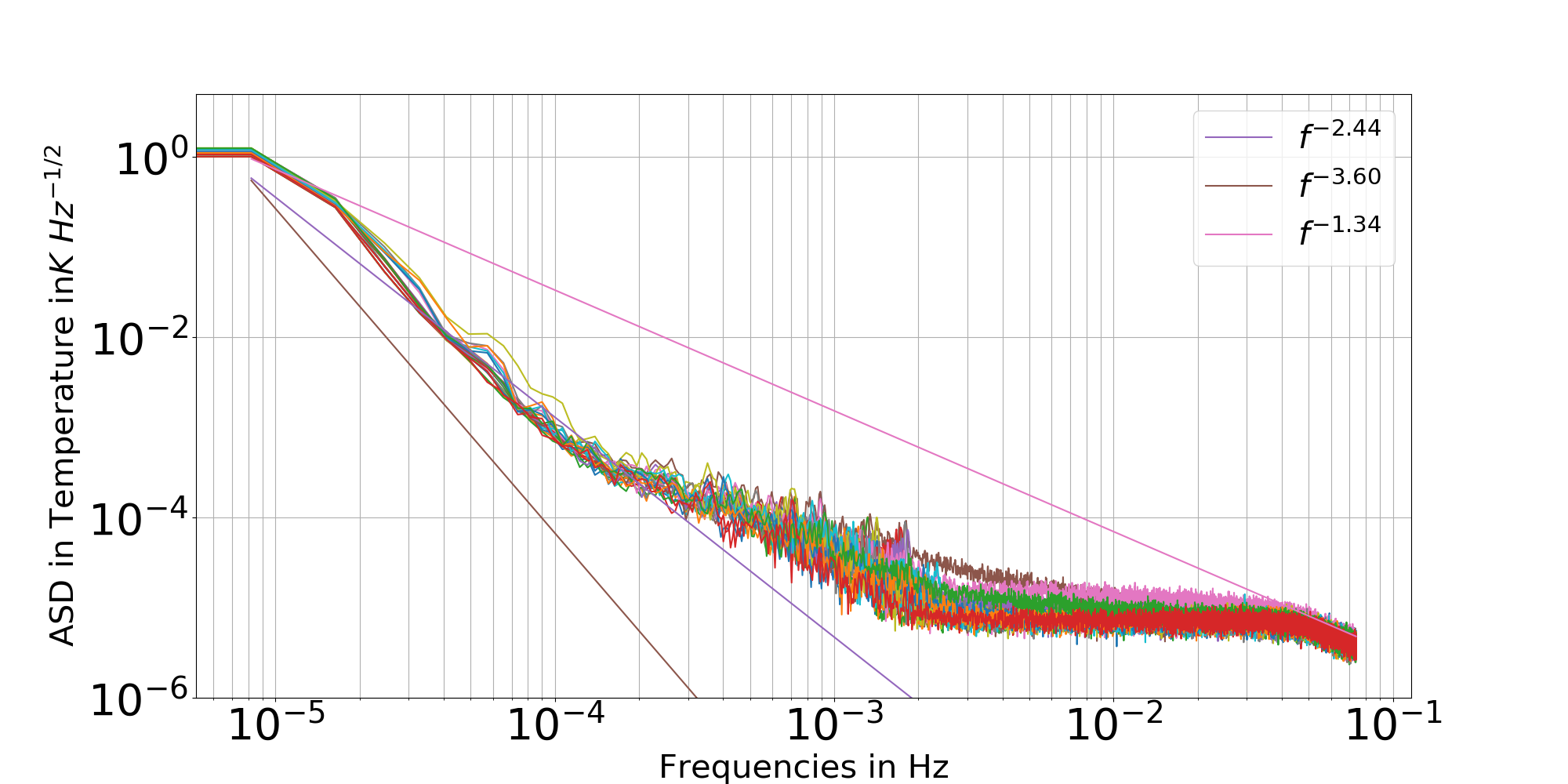}
    \caption{17-day averaging of the ASD of all thermal sensors (see Table~\ref{tab:thermal_sensor}) of the LPF satellite. }
    \label{fig:stack_ASD2}
\end{figure}

We separate the data into one-day segments. We average the periodograms over the 17 days of the thermal sensors to reduce fluctuations. The same process is used for the differential acceleration. Figure~\ref{fig:stack_ASD2} displays the ASD of the temperature sensors during the period of the month of February 2017 \cite{Armano:2019ekt}. 
Also displayed is the fit using the model of Armano {\it et al}.~\cite{Armano:2019ekt}.

\subsubsection{Coherence between temperature and acceleration}

\begin{figure}[htbp]
\centering
\includegraphics[width=\linewidth]{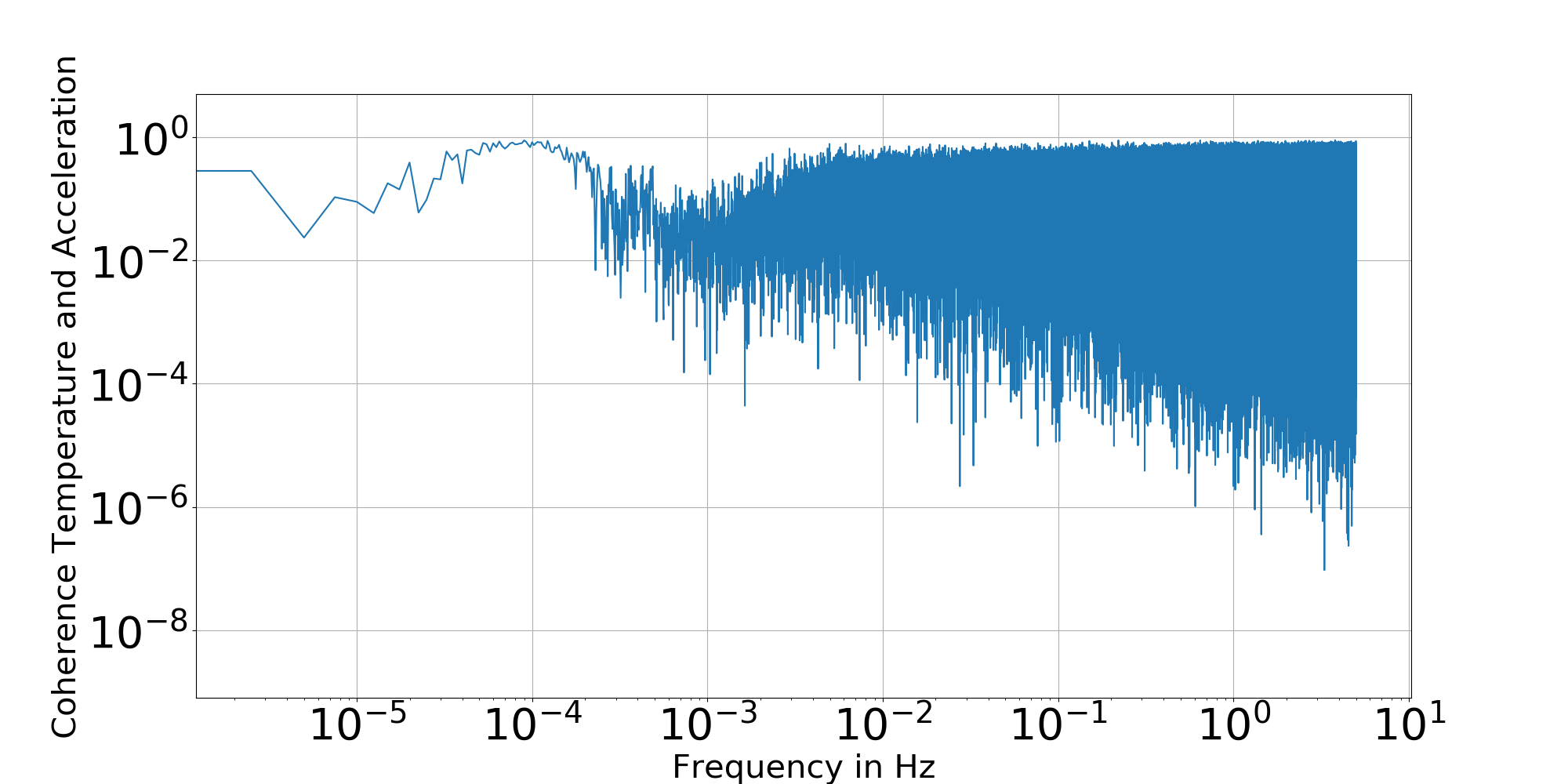}
\caption{Coherence of the mean temperature of the satellite and the differential acceleration of the test masses during the 17 days.}
\label{fig:coherence_tem}
\end{figure}

\begin{figure}[htbp]
\centering
\includegraphics[width=\linewidth]{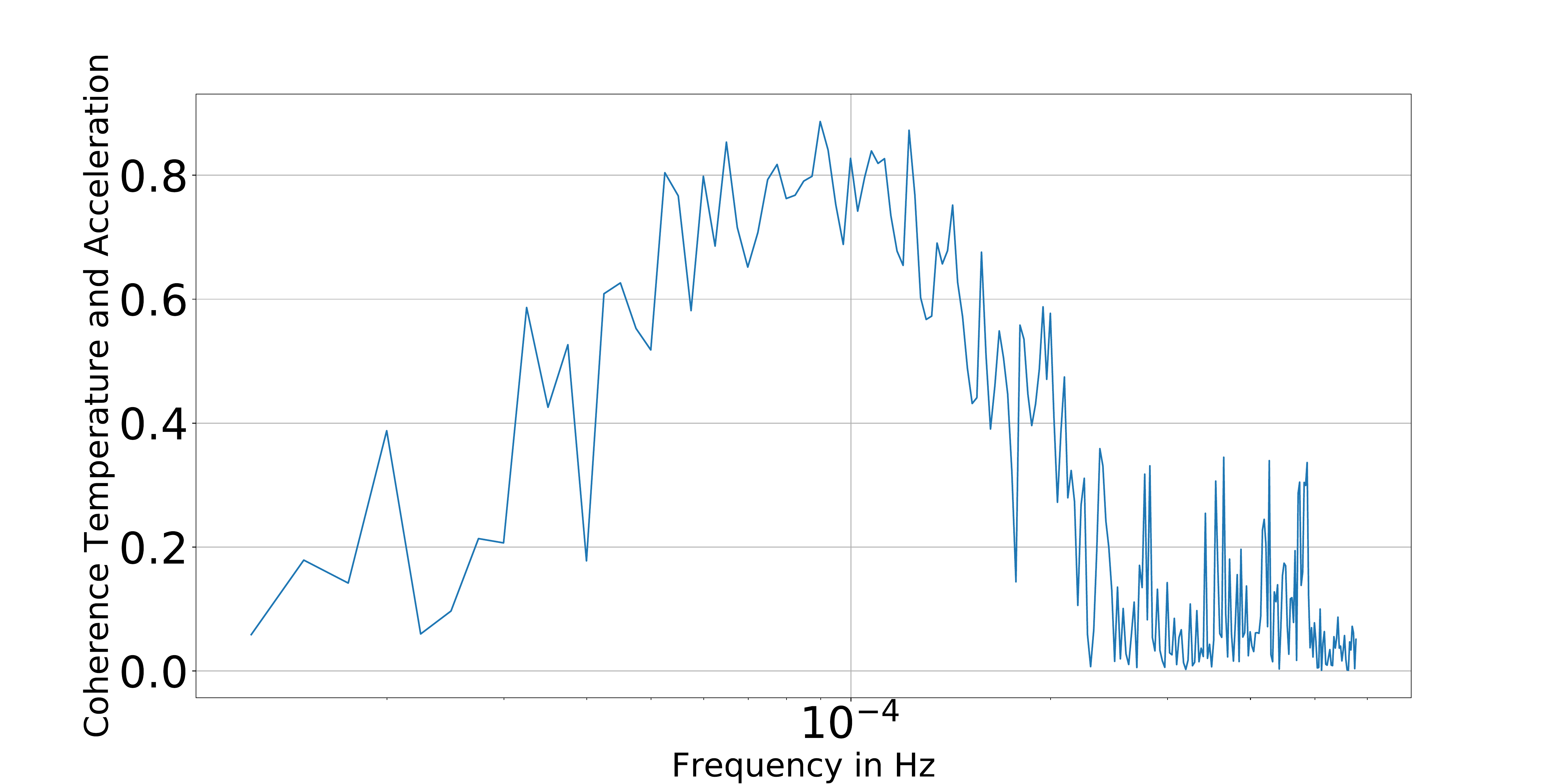}
\caption{Enlargement of the coherence of the mean temperature of the satellite and the differential acceleration of the test masses during the 17 days.}
\label{fig:coherence_tem_zoom}
\end{figure}

We present the coherence between the temperature and the differential acceleration in Fig.~\ref{fig:coherence_tem}. We average over the 17 days and over the 24 sensors to produce the estimation of the coherence between the thermal environment of the spacecraft fluctuation and the difference acceleration of the TMs.
We interpolate the thermal channels to change the data sample rate, but we cannot study the frequency above the sample rate of $ f_s = 0.2083 $ Hz. 
There is evidence of a correlation at $ f \simeq 1. \times 10 ^{- 4} $ Hz; we show this region in more detail in Fig.~\ref{fig:coherence_tem_zoom}.
The Pearson coefficient is $ 0.98$ for the frequency band [$1.\times10^{-5}$ Hz, $5.8 \times 10^{-4}$ Hz]. 

\subsection{Summary of the correlation measurements}
    \begin{figure}[htbp]
        \centering
        \includegraphics[width=\linewidth]{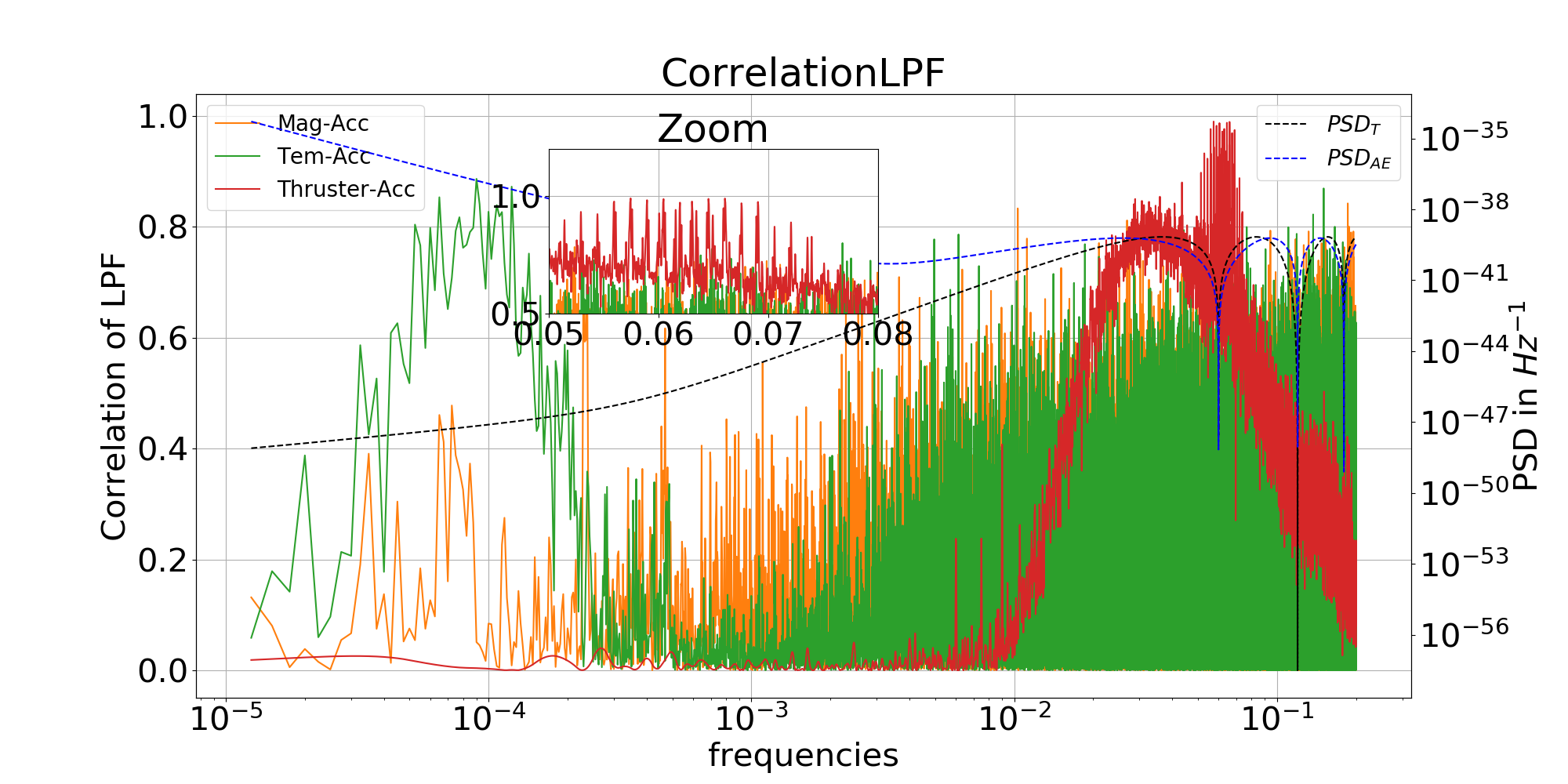}
        \caption{Coherence between the LPF test mass differential acceleration and the tested channels: magnetic field, temperature and $\mu$ thruster. The black and blue dotted lines are respectively the PSDs of the $T$ and $A = E$ channels of LISA. }
        \label{fig:LPFCorrinLISA/Corr}
    \end{figure}

In Fig.~\ref{fig:LPFCorrinLISA/Corr} we summarize the coherence measurements between the LPF differential acceleration and the magnetic field (red), temperature (green), and $\mu$ thrusters (orange). We also display the PSDs of the LISA channels $A$, $E$, and $T$ [see Eq.~(\ref{eq:modelPSDs})] to see in where the LPF correlations lie with respect to the LISA sensitivity. We want to understand which part of the LISA observational band would be affected by the addition of noise similar to what was observed in LPF.

\section{Addition of LPF Correlated Noise to the  LISA noise model}\label{sc:LPF+LISA}

\subsection{LISA noise model}
The amplitude of the LISA noise level budget  is given in the \href{https://atrium.in2p3.fr/nuxeo/nxdoc/default/f5a78d3e-9e19-47a5-aa11-51c81d370f5f/view_documents}{LISA science requirement document} and \cite{2019arXiv190706482B}. To create the data for our study, we use an acceleration noise of $ N_{acc} = 1.44 \times 10^{-48} \ \text{s}^{-4} \text{Hz}^{-1} $ and the fluctuation of the length of the optical path $ N_{Pos} = 3.6 \times 10^{-41} \ \text{Hz}^{-1}$. This corresponds to a data measurement period of four years. 

\begin{equation}\label{eq:modelPSDs}
\left\{
\begin{array}{l}
    PSD_A = N_A \\
    PSD_E = N_E \\
    PSD_T = N_T
\end{array}
\right.
\end{equation}
The noise components $ N_A (f) = N_E (f) $ and $ N_T (f) $ can be written as~\cite{PhysRevD.100.104055} 
\begin{equation}
\left\{
\begin{array}{l}\label{eq:lisamodel}
    N_A = N_1 - N_2 \\
    N_T = N_1 + 2 N_2 ~ , 
\end{array}
\right.
\end{equation}
with 
\begin{equation}\label{eq:lisamodel2}
\left\{
\begin{array}{l}
    N_1(f) = \left(4 S_s(f) + 8\left( 1 + \cos^2\left(\frac{f}{f_*}\right)\right) S_a(f)\right)|W(f)|^2 \\
    N_2(f) = -\left(2 S_s(f) + 8 S_a(f)\right)\cos\left(\frac{f}{f_*}\right)|W(f)|^2 ~ ,
\end{array}
\right.
\end{equation}
$W(f) = 1 - e^{-\frac{2if}{f_*}}$, and
\begin{equation}\label{eq6}
\left\{
\begin{array}{l}
    S_s(f) = N_{Pos} \\
    S_a(f) = \frac{N_{acc}}{(2 \pi f)^4}\left( 1 + \left(\frac{0.4 \text{ mHz}}{f} \right)^2 \right) ~ .
\end{array}
\right.
\end{equation}
Channels $A$, $E$, and $T$ are derived from the LISA TDI channels $X$, $Y$, and $Z$ \cite{Vallisneri_2012}.
We assume that the measurement of the gravitational wave signal by LISA is the same for all three channels, but with a phase difference of $ 2 \pi/3 $; this comes from the different orientations of each Michelson with unequal arms. It is also assumed that the noise is correlated between the channels $X,Y,Z $~\cite{Armstrong_1999,Vallisneri:2012np}; for example, a source of correlation between two channels is given by $N_2(f)$. Nominally, the $T$ channel does not contain a gravitational wave signal, but contains the uncorrelated LISA noise. This assumption is not perfectly correct, but for this analysis we will consider it to be true.
The relationships are:
\begin{equation}\label{eq:AET}
\left\{
\begin{array}{l}
A = \frac{1}{\sqrt{2}}(Z-X) \\
E = \frac{1}{\sqrt{6}}(X-2Y+Z) \\
T = \frac{1}{\sqrt{3}}(X+Y+Z) ~ .
\end{array}
\right.
\end{equation}
We assume that channels $A$ and $E$ contain the same gravitational wave information and the same noise level. The $A$ and $E$ noise parameters are modeled by the same parameters as the $T$ channel. 

\subsection{LPF correlations integrated into LISA noise parameters}\label{sc:modificationLISAnoise}

We extrapolate our observations of noise correlations in LPF and assume that the correlated noise is similar in  LISA. The architecture of the two spacecraft is somewhat comparable in terms of geometry and technology, although admittedly not identical. Some items, for example the $ \mu$ thrusters, will likely be the same. 
For our model the correlation noise in LPF will be an acceleration disturbance noise in LISA. We replace $N_{acc}$ in Eq.(~\ref{eq6}) by: 
\begin{equation}\label{eq:N_acc(f)}
    N_{acc,corr}(f) = N_{acc}\left(1+\frac{1}{M} \sum_{k=1}^M C_{LPF acc, n_k}(f)\right) ~ .
\end{equation}
with $C_{LPF acc, n_k}$ the coherence between the acceleration noise and the noise channel $n_k$ and $M$, the number of correlations considered~\cite{Coughlin_2016}. In our case, the $n_k$ are the $ \mu$ thrusters, the magnetic field and the temperature of the spacecraft. In a general case, it is also possible to add a correlated noise in the noise channels of LISA, with the update of the amplitude level of the optical path $ N_{pos}$.

\subsection{Correlations of LPF added with time delay interferometry}

The LISA constellation of three spacecraft will be placed in a heliocentric orbit and will form an equilateral triangle with arm lengths of 2.5 million kilometers. The distance from each satellite to another will be measured by laser beams. The constellation's orbit forms an angle of delay of 20° with respect to that of the Earth. Just as LISA performs a one-year period orbit, during this time it also performs a revolution on itself. For this study, we use the TDI version of the mock LISA data challenge \cite{Babak_2008}. In each satellite of the LISA constellation, there are two MOSAs, an optical fiber establishes the only optical link between the two MOSAs. Each spacecraft is in communication via a long arm with another satellite. So, we have three long arms and six MOSAs. Hence, we have six levels of positional noise and six magnitudes of acceleration. In each MOSA, we assume that the positional noise is a zero-mean Gaussian noise with a spectral density of $S_{p}(f) = N_{pos}$ and acceleration noise of $S_a(f) = \frac{N_{acc}}{(2 \ pi f)^4}\left( 1 + \left(\frac{0.4 \text{ mHz}}{f} \right)^2 \right)$. We assume that we have the same amplitude noise amplitude for each MOSA. Then we introduce a noise of different amplitude and study the noise deviations in the channels $ A $, $ E $ and $ T $. We also assume the LISA configuration with equal arms. 

\begin{figure*}[htbp]
    \centering
    \subfigure{\includegraphics[width=0.49\textwidth]{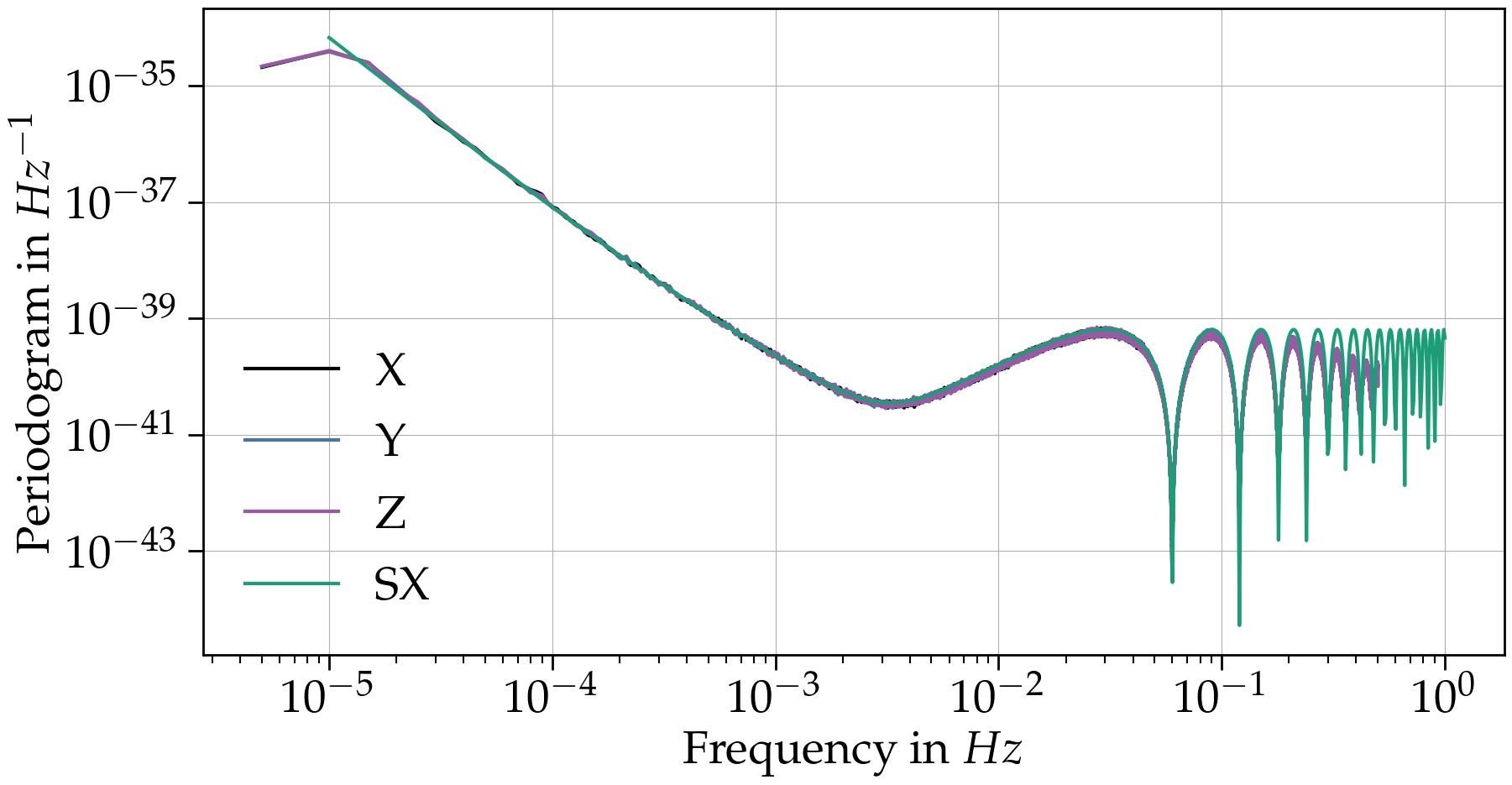} } 
    \subfigure{\includegraphics[width=0.49\textwidth]{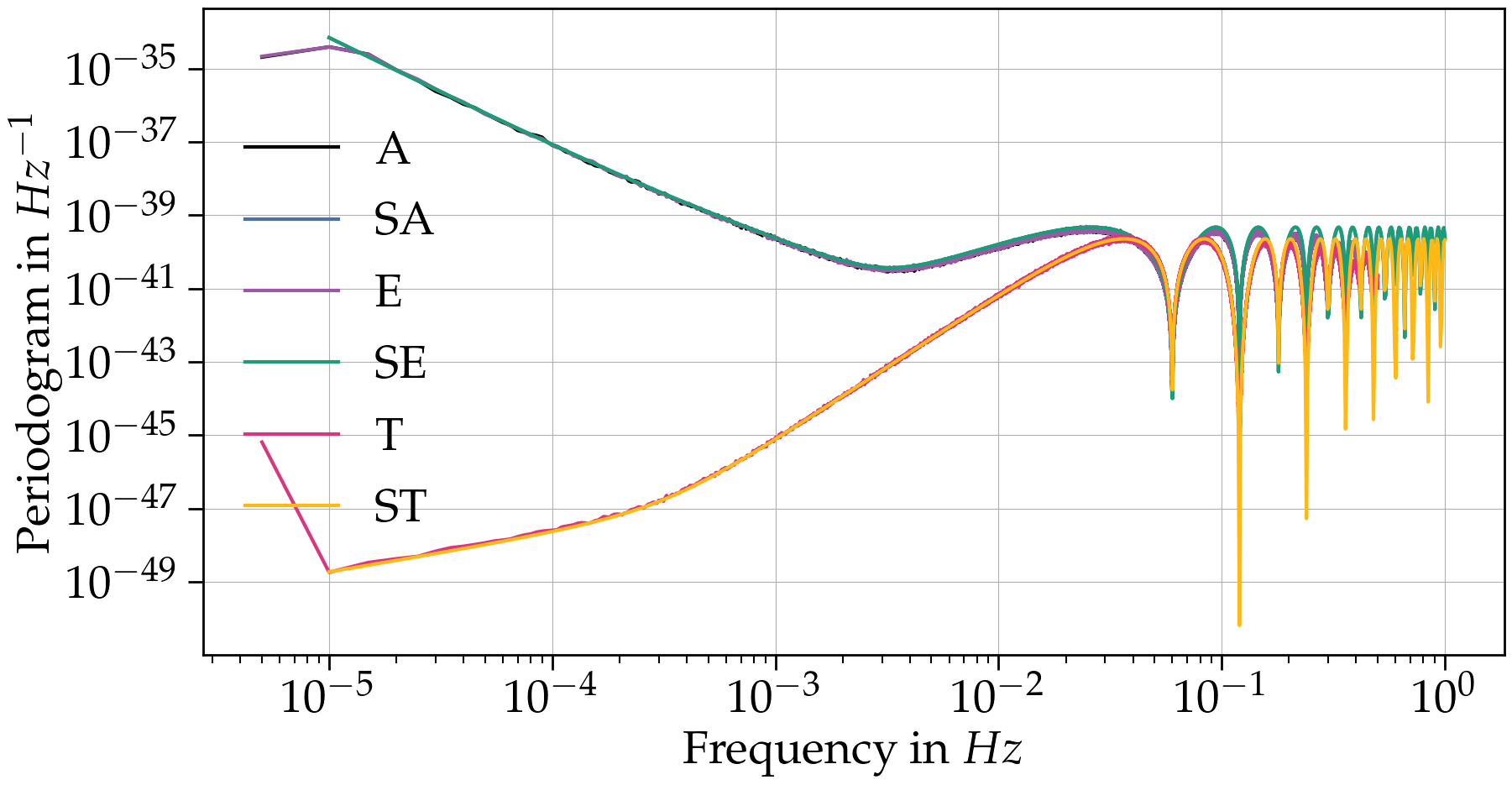} } 
\caption{Periodograms of channels $XYZ$ (a) and $AET$ (b) with LPF correlations and the LISA noise model [$S_X$, $S_A$, $S_E$, and $S_T$, see Eqs~(\ref{eq:lisamodel})~and~(\ref{eq:lisamodel2})]  of Smith and Caldwell \cite{PhysRevD.100.104055}  of channels $ XYZ $ and $ AET $.  }
\label{fig:LPFPSD}
\end{figure*}

We call $ n^p_{ij}(t) $ and $ n^a_{ij}(t)$, respectively the time series of the position and acceleration noise of the link connecting spacecrafts $i$ and $j$, in the direction $i$ toward $j$ (Gaussian noise following the power spectral density distribution of LISA noise see LISA noise budget \cite{LISAperfo}). We build the Doppler channels for each arm $ij$ and $ji$: 

\begin{equation}
h_{ij}(t) = n^p_{ij}(t) - n^a_{ij}(t) + n^a_{ji}(t-L)
\end{equation}
\begin{equation}
h_{ji}(t) = n^p_{ji}(t) + n^a_{ji}(t) - n^a_{ij}(t-L) ~ .
\end{equation}
We form the Michelson channels: 
\begin{equation}
h_{i}(t) = h_{ij}(t-L) - h_{ji}(t) + h_{ik}(t-L) - h_{ki}(t) ~ .
\end{equation}
With the Michelson channels, we can form the TDI channels $ X, Y, Z $ by subtracting a delayed Michelson signal of $ L $ (with $ c = 1 $) with a nondelayed one, 
\begin{equation}
U(t) = h_{i}(t) - h_{i}(t-L) ~ ,
\end{equation}
with $U = X,Y,Z$. So, we finally have the TDI $ X, Y, Z $ channels, with which we can form the $ A, E, T $ channels with Eq.~(\ref{eq:AET}). If we add the correlation to the means described in Sec.~\ref{sc:modificationLISAnoise}, we can inject the LPF correlations into LISA's TDI channels. 
Figigure \ref{fig:LPFPSD} shows periodograms of the resulting TDI calculation with the addition of the LPF correlations at the levels observed with our previous studies of the LPF data. We add the correlations as a modification of the noise level of acceleration amplitude [see Eq~(\ref{eq:N_acc(f)})]. Note in Fig.~\ref{fig:LPFPSD} the difference at the low frequencies of the periodogram. This is where the measurement of the SGWB by LISA is most important. 

\section{Spectral separation of SGWB in the context of adding LPF correlations in LISA}\label{sc:spectralseparatin}

 We reproduce the spectral separation study for LISA to describe a SGWB, as described in~\cite{PhysRevD.103.103529}, but now with two scenarios: with and without LPF-like correlations, to see the effect on measuring the cosmologically produced SGWB. There are several techniques to separate the SGWB signal from the LISA noise. 
An approach based on  principal component analysis was proposed by M. Pieroni and E. Barausse \cite{Pieroni_2020} to demonstrate the extraction of the cosmological SGWB and astrophysical foreground with LISA noise. This method is robust and can be extended to different detectors.
A spectral shape reconstruction based on a binning procedure~\cite{Caprini:2019pxz}  demonstrated the possible reconstruction of different functional forms of the SGWB and their separation from the LISA noise. The method is agnostic of signal and spectral shape. It requires stationary noise over a long period of time. The reconstruction of the spectral shape can be expanded to take  unequal arm lengths of the LISA constellation into account~\cite{1818908}.
A fast excess power approach was used to extract a stationary and isotropic Gaussian SGWB from LISA noise ~\cite{10.1088/1361-6382/abb637}.

To provide an estimate of the uncertainty of the amplitude of the cosmological SGWB, we calculate the {\em coefficient of variation}, i.e. the ratio of the standard deviation to the mean of the estimate.  The standard deviation and mean based on an adaptive Markov chain Monte Carlo (A-MCMC) analysis are obtained by taking the posterior sample standard deviation and mean of the MCMC chain, respectively. 
 For  standard error estimates based on the Fisher information, we use for each input parameter the square root of the diagonal of the covariance matrix. The coefficient of variation is usually expressed as a percentage. 
 It is then easy to compare  two uncertainty estimates. A good overlap indicates that the studies are consistent.

Our goal is to demonstrate the ability for LISA to measure a cosmological SGWB in the presence of other stochastic signals (an astrophysically produced SGWB, standard LISA noise) and correlated acceleration noises in the test mass system.  
The SGWB is typically quantified by its normalized energy density as a function of frequency, $\Omega_{GW} (f)$, namely the energy density $d\rho_{GW}$ of GWs per logarithmic frequency interval $d \ln(f)$, divided by the critical energy density of the Universe  $\rho_c = 3H_0^2c^2/(8\pi G)$, $c$ the speed of light, $G$ is Newton's constant, and  $H_0$ is the Hubble constant,:
\begin{equation}
\label{eq:energyDensity}
\Omega_{GW}(f) = \frac{1}{\rho_c}\frac{d\rho_{GW}}{d \ln(f)} ~.
\end{equation}
We use the  estimation from Chen \textit{et al.} 
for an astrophysical SGWB that we inject into the data, $\Omega_{Astro}(f) = \Omega_{2/3}\left(\frac{f}{f_*}\right)^{\alpha_{2/3}}$, with $\alpha_{2/3} = \frac{2}{3}$ and $\Omega_{2/3} = 4.4 \times10^{-12}$ at $f_* = 3$ mHz~\cite{2019ApJ...871...97C}. This result is based on  the LIGO-Virgo observations of binary black hole and binary neutron star mergers from the O2 observing run, $\Omega_{GW}(25~\text{Hz}) = 8.9^{+12.6}_{-5.6} \times10^{-10}$~\cite{LIGOScientific:2019vic}. The most recent upper limit given by LIGO-Virgo-KAGRA from the O3 observing run is $\Omega_{GW}(25~\text{Hz}) = 7.2^{+3.3}_{-2.3} \times10^{-10}$~\cite{PhysRevD.104.022004}. The level from Chen \textit{et al.}~\cite{2019ApJ...871...97C} corresponds to the upper level from the LIGO-Virgo O2 results, and is a conservative choice for our study.

We fit the level of the cosmological SGWB to determine its detectability, where  $\Omega_{Cosmo}(f) = \Omega_0\left(\frac{f}{f_*}\right)^{\alpha_0}$, with $\alpha_0 = 0$, hence $\Omega_{Cosmo}(f) = \Omega_0$.
We have, from our previous study \cite{PhysRevD.103.103529}, presented the evidence of the separability of cosmological and astrophysical backgrounds with a LISA observation threshold around $\Omega_{0} \approx 1 \times 10^{-12} $ to $1 \times10^{-13} $. 

We calculate the measurement uncertainty of the cosmological SGWB amplitude for normalized cosmological energy densities $\Omega_{0}$ between $ 1 \times 10^{-14} $ and $ 1 \times 10^{-8} $. We set the limit of detectability when there is less than 20\% uncertainty on the amplitude of the normalized cosmological amplitude of the energy spectral density of the cosmologically produced SGWB. This follows the widely used rule of thumb that a coefficient of variation below 20\% is considered as good, as e.g.\ in assessing the variability in  agricultural experiments \cite{Ferreira2016}, quantitative assays in clinical chemistry \cite{Reed2002}, and magnetic resonance imaging \cite{Aronhime}.

We have previously made Bayesian studies where we use an A-MCMC algorithm~\cite{PhysRevD.103.103529,2021arXiv210504283B,Boileau:2021gbr} to estimate the parameters of our model. Here we will estimate two parameters for the LISA noise, two parameters for the cosmological background (amplitude and slope), and similarly two parameters for the astrophysical background. In total, we fit six parameters on the two periodograms for channels $ A $ and $ E $, and simultaneously, the two noise parameters with the channel $ T $. We assume that the effect of the LPF correlations will be an increase in the amplitude of the LISA acceleration noise.  
The LPF correlations are therefore added in LISA as a noise correlated to the set of MOSAs of the LISA constellation. 

\begin{figure}[htbp]
\includegraphics[width=\linewidth]{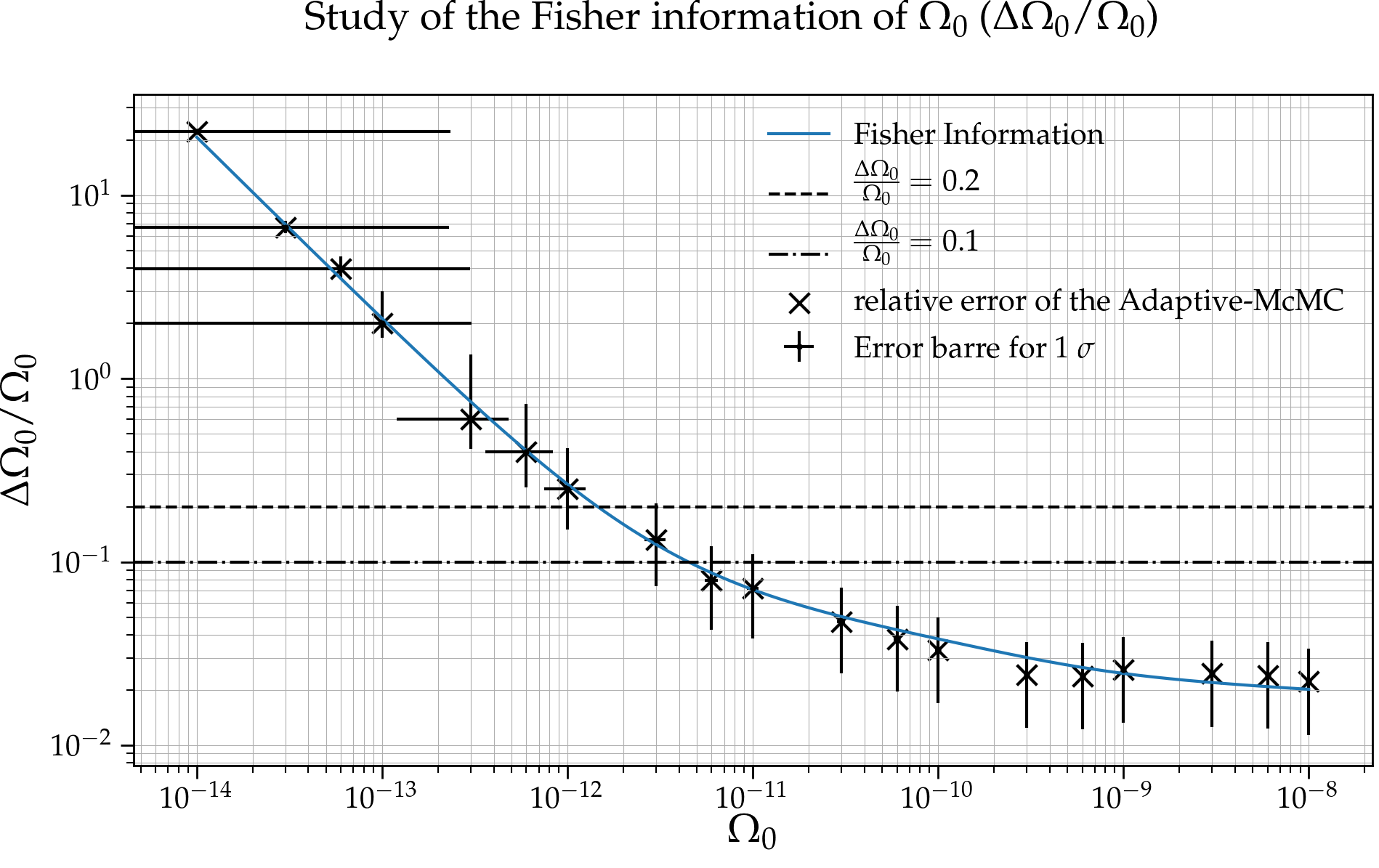}
\caption{Estimation of the uncertainty of the cosmological SGWB amplitude, with the study of the Fisher information in blue and the A-MCMC represented by the black scatter. The upper horizontal dashed line represents the error level at $ 20 \% $ (limitation criterion). }
\label{fig:Uncertaintylpf}
\end{figure}

\begin{figure}[htbp]
\includegraphics[width=\linewidth]{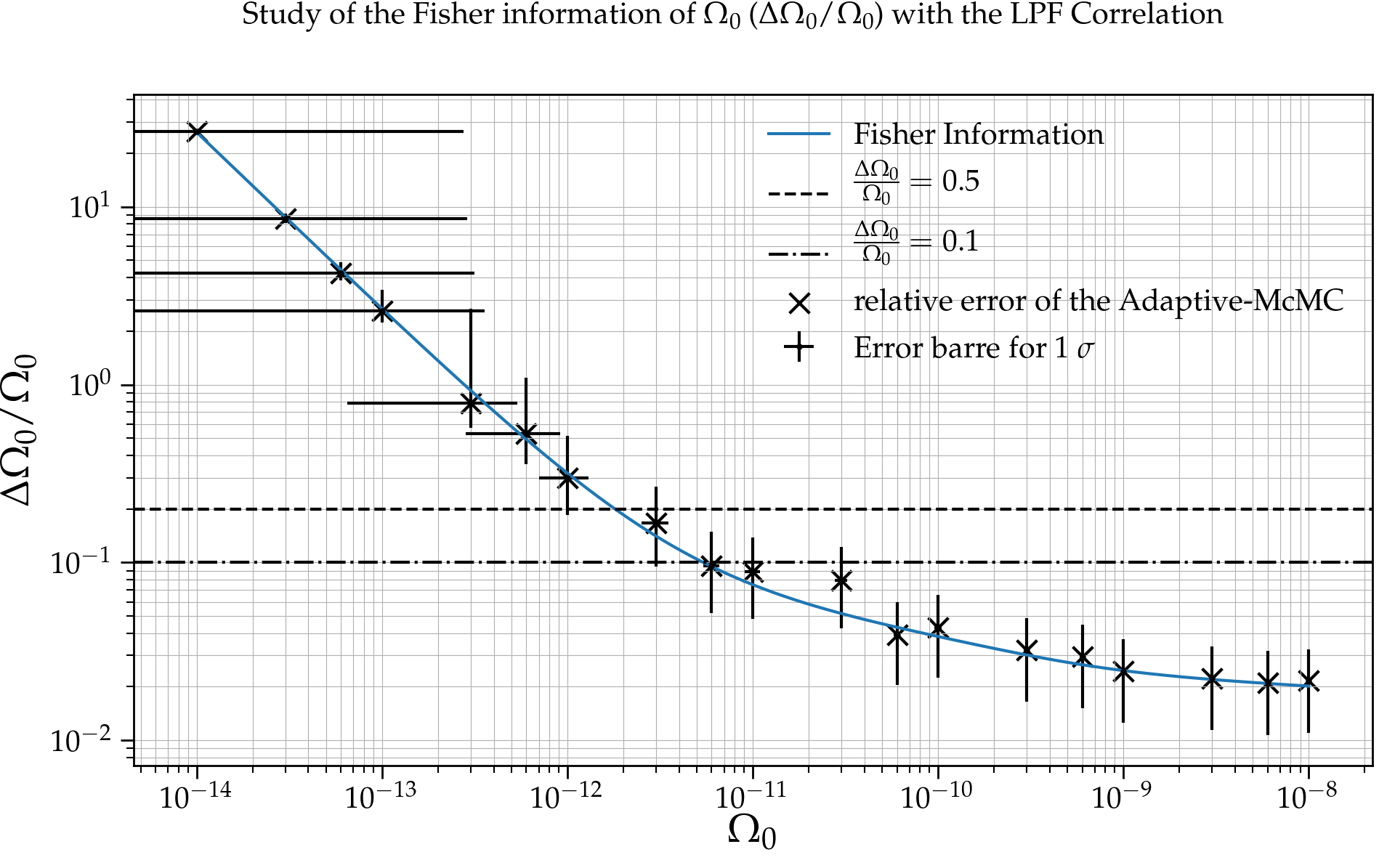}
\caption{Estimation of the uncertainty of the cosmological SGWB amplitude in the case of adding the correlations of LPF, with the Fisher information study in blue and the A-MCMC represented by the black scatter. The upper horizontal dashed line represents the error level at $ 20 \% $ (limitation criterion).  }
\label{fig:Uncertaintycorr}
\end{figure}

Figs~\ref{fig:Uncertaintylpf} and \ref{fig:Uncertaintycorr} are, respectively, the estimate of the uncertainty of the cosmological SGWB amplitude using the Fisher information (blue line) and from the A-MCMC (black scatter) with and without LPF correlation. The results are presented in Table~\ref{tab:Uncertainty}. LPF correlations introduce a small performance loss with our spectral separation method, namely a loss of 22 \% for the cosmological SGWB amplitude.

\begin{table}[htbp]
\centering
\begin{tabular}{|c|c|}
\hline
& $\Omega_{0}$ \\ \hline
Limit without LPF & $1.4 \ \times 10^{-12}$ \\ \hline
Limit with LPF  & $1.8 \ \times 10^{-12}$ \\ \hline
\end{tabular}
\caption{Limitation of SGWB cosmological amplitude measurement with and without LPF correlations. The cosmological SGWB is $\Omega_{Cosmo}(f) = \Omega_0$, while the astrophysical SGWB is $\Omega_{Astro}(f) = \Omega_{2/3}\left(\frac{f}{f_*}\right)^{2/3}$, with  $\Omega_{2/3} = 4.4 \times10^{-12}$ \cite{2019ApJ...871...97C}.}
\label{tab:Uncertainty}
\end{table}

\section{Injection and modification of the LISA noise}\label{sc:ModLISA}
In this section, we further study the effect of LISA noise modifications on the SGWB measurement. Specifically, we explore the possibility of correlated noise between the two optical systems on a particular LISA satellite. 

\subsection{Correlated noise and spectral separation of the stochastic background}

Here we examine a LISA SGWB search in the context of correlated noise. To do this, we introduce a correlated noise into our implementation of the LISA TDI. The LISA noise correlation sources can be multiple. For example, one could also add interferometric noise. It is not the goal here to make a zoology of correlations, but simply to understand the behavior of adding  correlation and observing its effect on the measurement of the SGWB. Hence for simplicity in this study we  add a correlation only to the acceleration noise between two MOSAs. In light of the LISA~\cite{lisacorr} working document, we observe with the LPF data a correlation between the temperature and the difference in acceleration at very low frequencies. Thus, we can hypothesize that the change in temperature of the satellite disturbs the free fall measurements of the test masses. Each satellite contains two MOSAs; these are the optical systems in which interferometric measurements are made. 

We can induce between two MOSAs of the same satellite a noise term correlated to the noise of acceleration coming from a variation of the temperature of the satellite. To do this, we introduce a noise term common to the two MOSAs of satellite 1, then we calculate the time series of the channels $X,Y,Z$. In this study, we are not trying to establish the value of the correlation with precision, but rather we are trying to understand the effect of a correlation between two MOSAs of the same satellite on the LISA model. In addition, we also seek to accurately measure the effect of the correlation and its impact on our cosmological SGWB measurements. We use the noise value of the temperature stability of the GRS (gravitational reference sensor) and the EH (electrode housing), See Sec.~1.3.6 in \cite{boileauPhD}. Indeed, the result is a force on the test-mass such that 

\begin{equation}\label{eq:GRS}
\begin{aligned}
   & S_g^{T_{GRS}} = \left| \frac{\partial g_x}{\partial T_{GRS}}\right|S_{T_{GRS}} \\ 
    &= 144 \text{ fm}^2 \text{s}^{-4}\text{Hz}^{-1}\left(\frac{\frac{\partial g_x}{\partial T_{GRS}}}{1\text{ pm}\text{s}^{-2}\text{K}^{-1}} \right)^2 \frac{S_{T_{GRS}}}{144\text{ mK}^2 \text{Hz}^{-1}} ~.
\end{aligned}
\end{equation}
As stated in the LISA~\cite{LISAperfo} noise budget document, we use the value from LPF mission $\frac{\partial g_x}{\partial T_{GRS}}=1\text{ pm}\text{s}^{-2}\text{K}^{-1}$. Additionally,  $S_{T_{GRS}} < 1.4 \text{ mK}\text{Hz}^{-1} \left(1+ \left(\frac{2\times10^{-3}}{f}\right)^4\right)$.  The limiting case of correlated noise of $S_g^{T_{GRS}}= (4\times10^{-12}\text{ m} \text{s}^{-2}\text{Hz}^{-1/2})^2 \left(1+ \left(\frac{2\times10^{-3}\text{Hz}}{f}\right)^4\right)$, this is a potentially overestimated case, but makes it possible to understand the effects at stake here on the measurement of the SGWB. Indeed, we voluntarily forced this noise to see an effect; it corresponds to the maximum temperature noise within a satellite. 

\begin{figure}[htbp]
    \centering
    \includegraphics[width=0.49\textwidth]{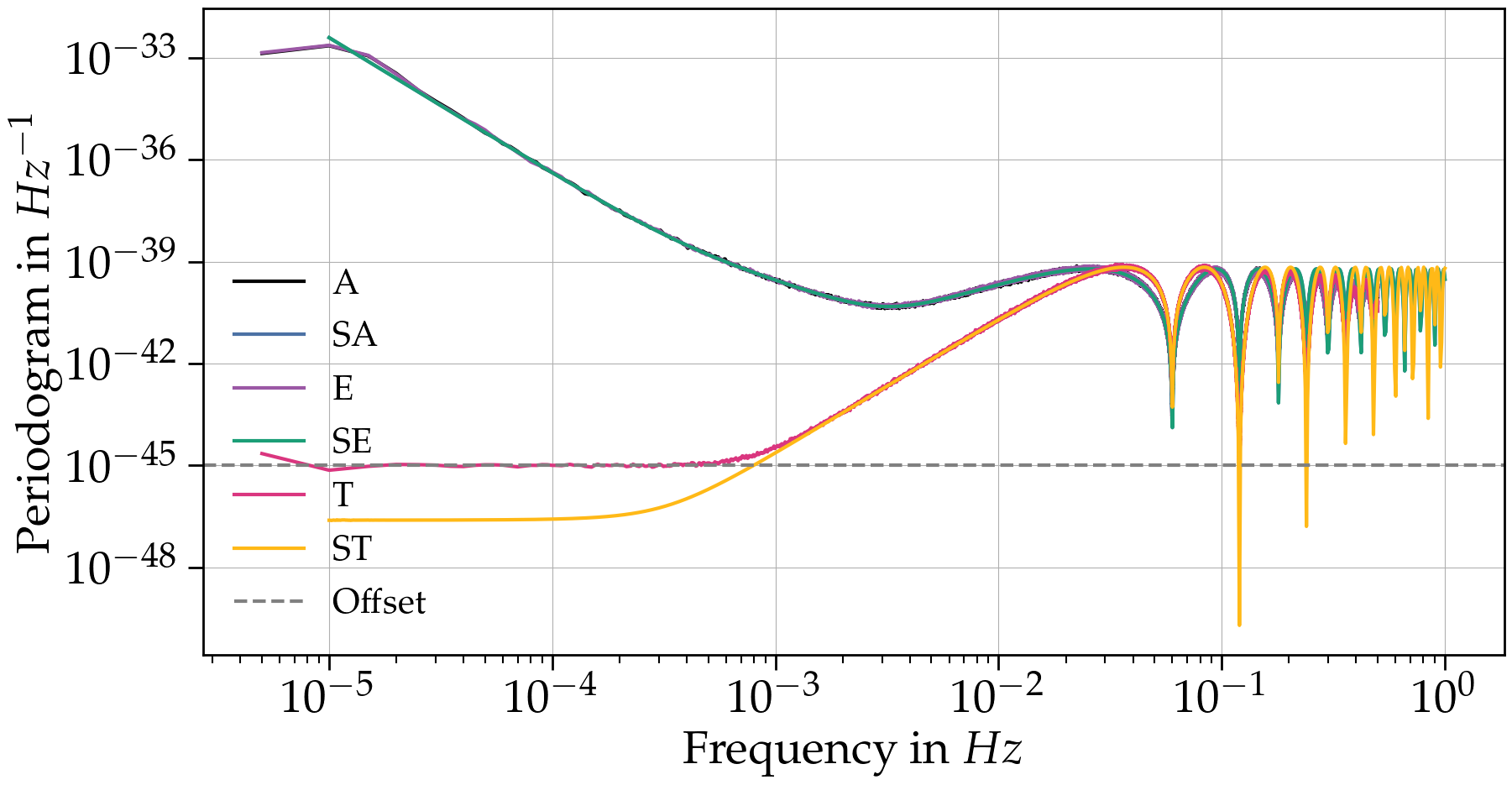}
\caption{Periodograms of channels $AET$ with additional GRS noise (see Eq.~\ref{eq:GRS}) and the LISA noise model ($S_A$, $S_E$ and $S_T$ [see Eqs.~(\ref{eq:lisamodel})~and~(\ref{eq:lisamodel2})] of Smith and Caldwell \cite{PhysRevD.100.104055}  of channels $ XYZ $ and $ AET $. The "Offset" dash gray line is the constant we need to add to the $T$ channel model to match the $T$ channel periodogram coming from the TDI calculation.  }
\label{fig:OffsetPSDAET}
\end{figure}

\begin{figure*}[htbp]
\includegraphics[width=\linewidth]{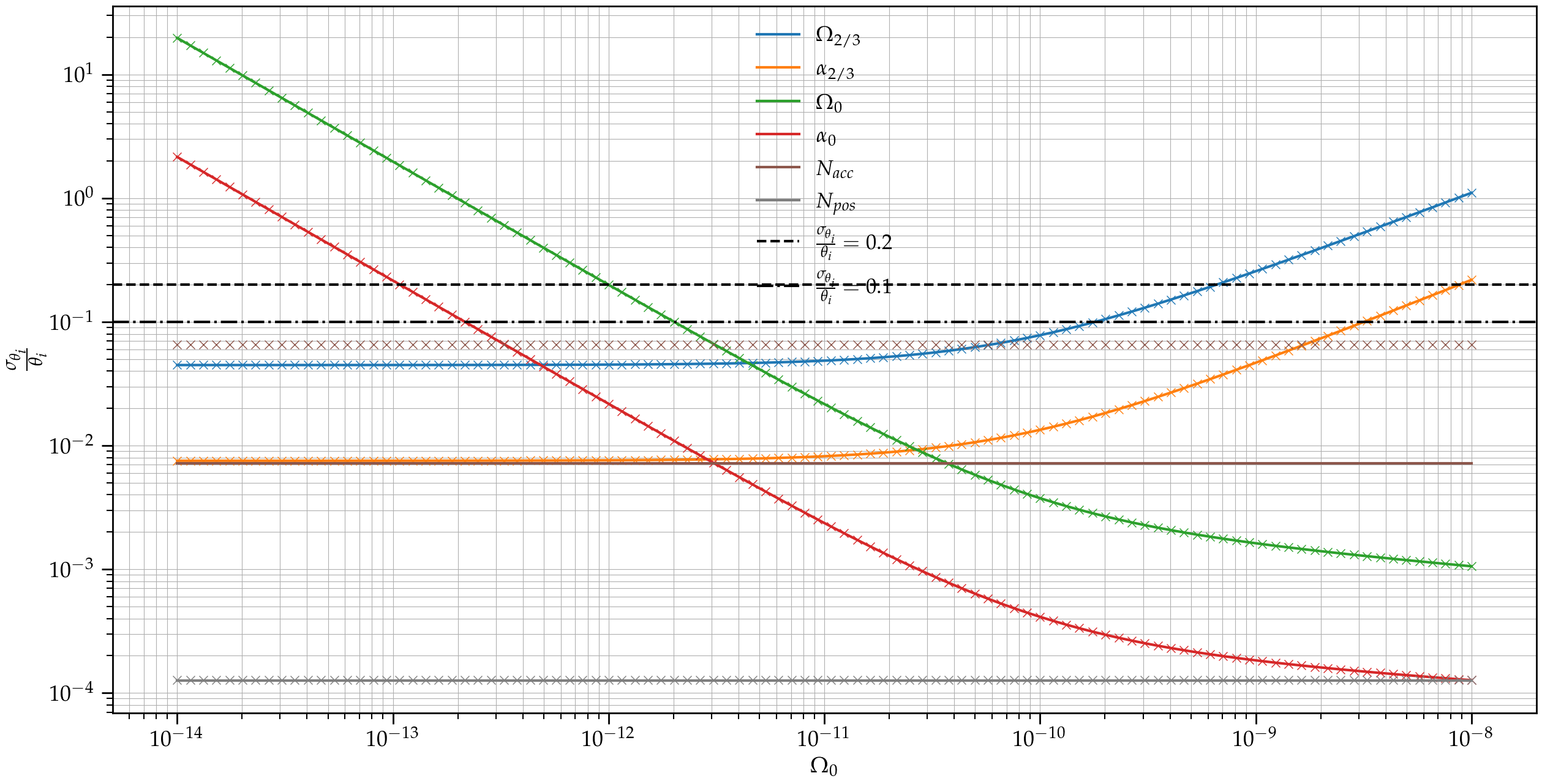}
\caption{Estimation of the uncertainty of LISA noise parameters ($N_{acc}, N_{pos}$), SGWB parameters ($\Omega_{2/3}$, $\alpha_{2/3}$, $\Omega_{0}$, $\alpha_{0}$)  with (scatter lines) and without (solid lines) the offset on the channel $T$. The upper horizontal dashed line represents the error level at $ 20 \% $ (limitation criterion).  }
\label{fig:Uncertaintyoffset}
\end{figure*}

We note by comparing periodograms from the TDI calculation with the addition of the GRS noise with the LISA noise model of Smith and Caldwell \cite{PhysRevD.100.104055} (see Fig.~\ref{fig:OffsetPSDAET}), a gain at low frequencies on the channel $T$. This one appears to be able to be modeled by the addition of a constant (offset) in the noise model of the channel $T$ .A fit, made with an MCMC on the PSD of the channel $T$, gives a value of \textit{Offset} $= 1 \times 10^{-45} \ \text{Hz}^{-1}$. Figure~\ref{fig:Uncertaintyoffset} shows the estimation of the uncertainty from the Fisher information study of the LISA noise parameters ($N_{acc}, N_{pos}$) and  SGWB parameters ($\Omega_{2/3}$, $\alpha_{2/3}$, $\Omega_{0}$, $\alpha_{0}$) with (scatter lines) and without (solid lines) the offset on the channel $T$. 

According to this study, if a correlated noise were to affect only the $T$ channel, the measurement of the SGWB does not seem to be affected. It is important to notice the possibility to accurately measure the effect of adding a correlation on the noise parameters of LISA. Our work here makes it possible to build a LISA figure of merit.

\subsection{LISA noise modifications}

In our two previous studies of the SGWB~\cite{PhysRevD.103.103529,2021arXiv210504283B}, we have considered different amplitudes for the astrophysical background ($\Omega_{astro,GW}(f) = \Omega_{2/3}\left(\frac{f}{f_{ref}}\right)^{2/3}$, with $f_{ref} = 25$ Hz) in the range $\Omega_{2/3} = 3.55 \times 10^{-10}$ to $3.55 \times 10^{-8}$. We have estimates for each astrophysical background, and using them, we determine what is the cosmological SGWB amplitude that LISA can measure \cite{PhysRevD.103.103529}. In the second study \cite{2021arXiv210504283B}, we have added a galactic foreground from white dwarf binaries, using the catalog of Lamberts {\it et al.}~\cite{10.1093/mnras/stz2834}. 
We have tested the influence of the modification of the LISA cosmological SGWB measurement with the Fisher information study. In these studies we have demonstrated the possibility of separating a cosmological background in the presence of an astrophysical background and a modulated galactic foreground. We have highlighted that the loss of cosmological background measurement performance was mainly related to the astrophysical background and LISA noise. We have produced similar results for when LISA searches for a cosmic string produced SGWB in the presence of a galactic foreground, an astrophysical background, and LISA noise~\cite{Boileau:2021gbr}. Because we observed that the astrophysical background and LISA noise produced the dominant effects, for simplicity we have not included the galactic foreground in this present study.

We note that the prediction of LISA noise is uncertain. Even though very good models exist now for predicting the LISA noise~\cite{LISAperfo}, other effects may come into play.
For example, micrometeoroids could deteriorate the LISA optical system and induce scattered light~\cite{10.1117/1.JATIS.6.4.048005}. To test the influence of a modification of the LISA noise, we estimate the uncertainty of the parameters of LISA noise parameters ($N_{acc}, N_{pos}$), SGWB parameters ($\Omega_{2/3}$, $\alpha_{2/3}$, $\Omega_{0}$, $\alpha_{0}$) in a context of modification of LISA noise parameters. 

\begin{figure*}[htbp]
\includegraphics[width=\linewidth]{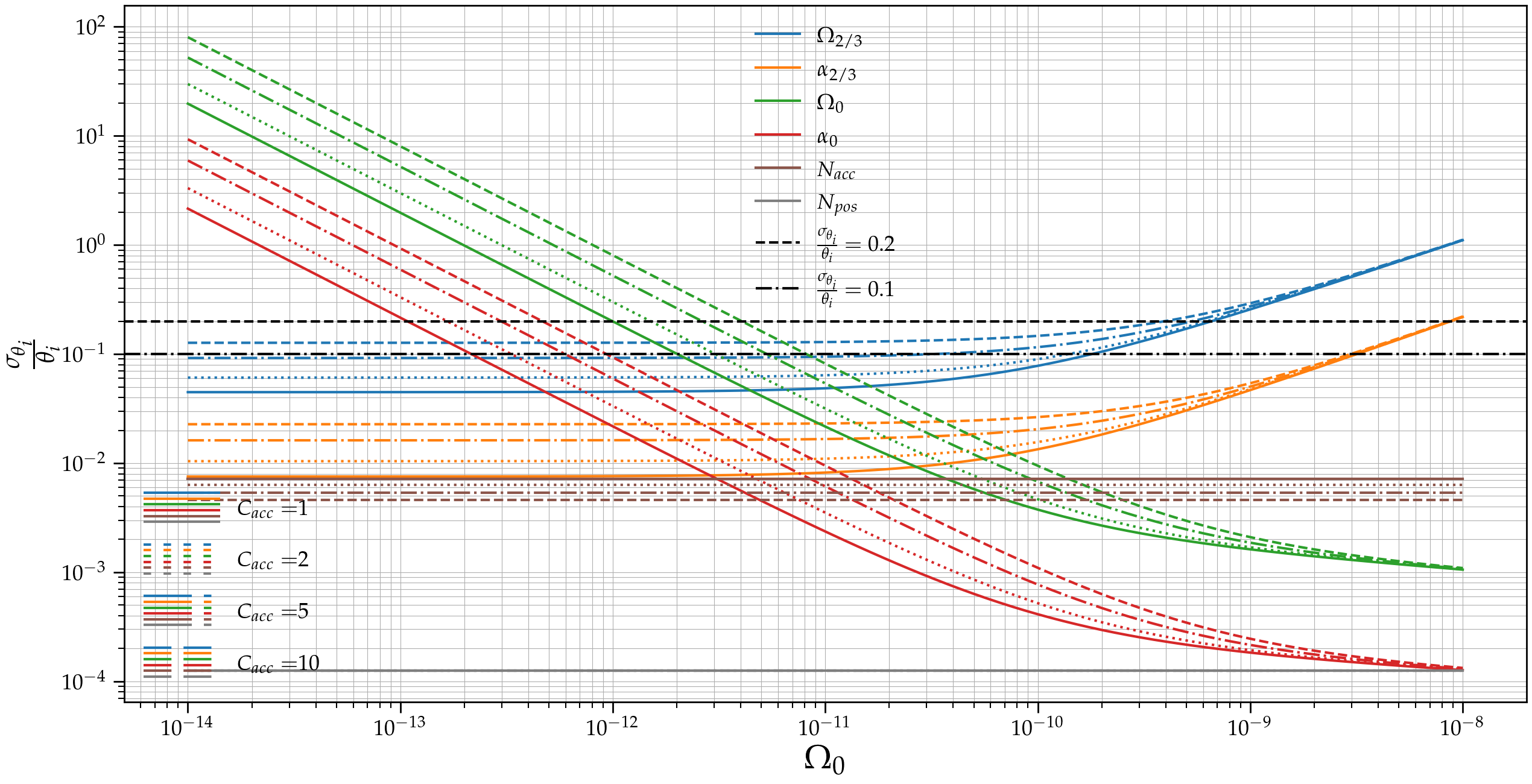}
\caption{Estimation of the uncertainty of LISA noise parameters ($N_{acc}, N_{pos}$), SGWB parameters ($\Omega_{2/3}$, $\alpha_{2/3}$, $\Omega_{0}$, $\alpha_{0}$)  with modifications of the acceleration noise. The upper horizontal dashed line represents the error level at $ 20 \% $ (limitation criterion).}
\label{fig:Uncertaintyacc}
\end{figure*}

\begin{figure*}[htbp]
\includegraphics[width=\linewidth]{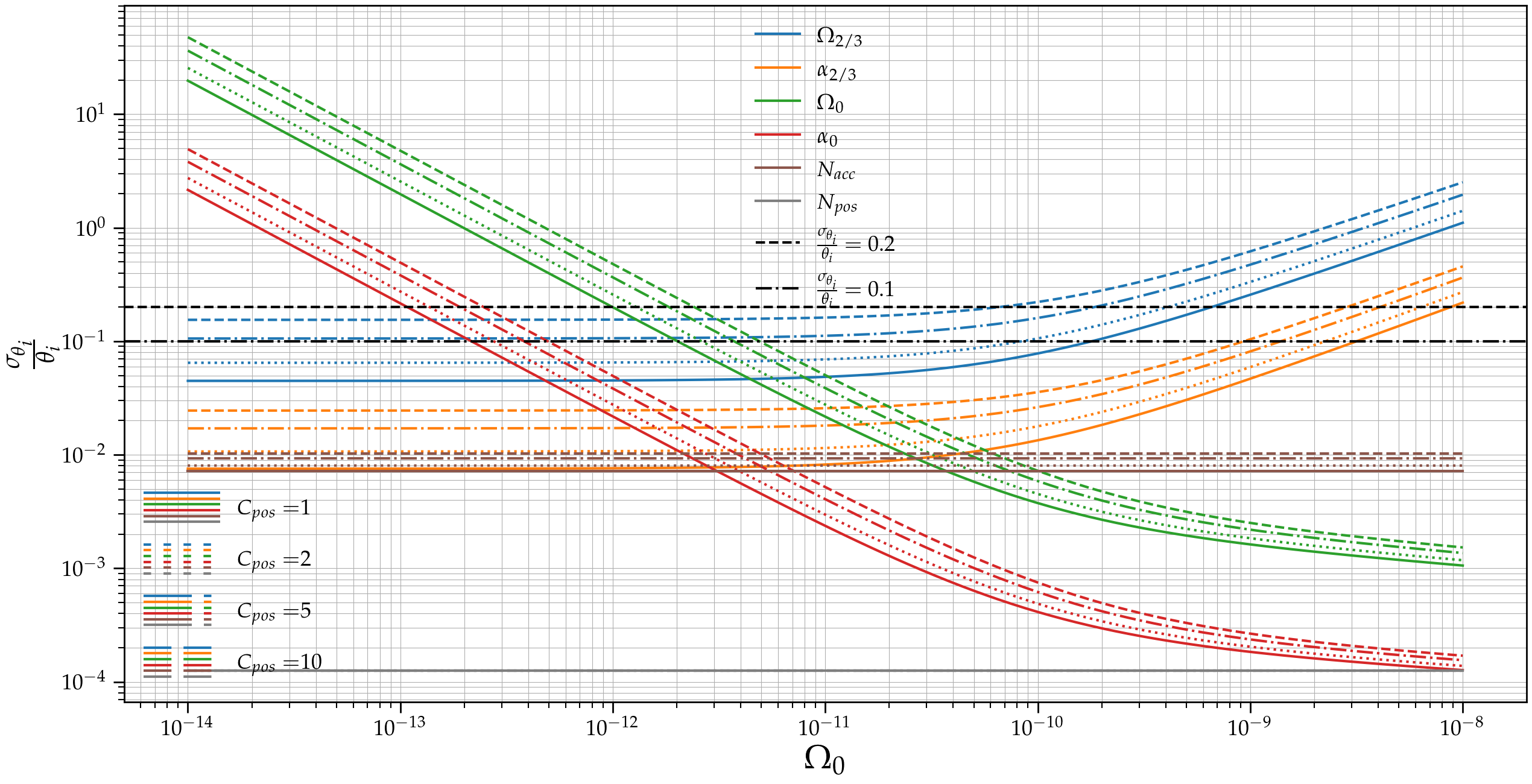}
\caption{Estimation of the uncertainty of LISA noise parameters ($N_{acc}, N_{pos}$), SGWB parameters ($\Omega_{2/3}$, $\alpha_{2/3}$, $\Omega_{0}$, $\alpha_{0}$)  with modifications of the optical noise. The upper horizontal dashed line represents the error level at $ 20 \% $ (limitation criterion).  }
\label{fig:Uncertaintypos}
\end{figure*}

We introduce a modification of the acceleration noise amplitude $N_{acc}$ by testing the separability for new forced value of $N_{acc}$, such as $N_{acc,new} = C_{acc} N_{acc}$, with $C_{acc} = [1,2,5,10]$. Figure~\ref{fig:Uncertaintyacc} shows the uncertainty of the different LISA noise parameters ($N_{acc}, N_{pos}$), and SGWB parameters ($\Omega_{2/3}$, $\alpha_{2/3}$, $\Omega_{0}$, $\alpha_{0}$). Respectively, Fig.~\ref{fig:Uncertaintypos} presents the uncertainty from the Fisher information with the modification of the optical noise level ($N_{pos,new} = C_{pos} N_{pos}$, with $C_{pos} = [1,2,5,10]$). We note that a degradation of the LISA noise decreases the probability to measure SGWB parameters. Limitation measurement estimations of the amplitude of the cosmological SGWB are given in the Table~\ref{tab:Uncertaintyaccpos}. The limitation criterion is given by the error level of $20 \%$.

\begin{table}[htbp]
\begin{tabular}{ cc } 
\begin{tabular}{|c|c|}
\hline
$C_{acc}$& $\Omega_{0,lim}$ \\ \hline
1 & $1.0 \ \times 10^{-12}$ \\ 
2  & $1.5 \ \times 10^{-12}$ \\
5  & $2.6 \ \times 10^{-12}$ \\
10  & $4.0 \ \times 10^{-12}$ \\ \hline
\end{tabular} & 
\begin{tabular}{|c|c|}
\hline
$C_{pos}$& $\Omega_{0,lim}$ \\ \hline
1 & $1.0 \ \times 10^{-12}$ \\ 
2  & $1.3 \ \times 10^{-12}$ \\
5  & $1.8 \ \times 10^{-12}$ \\
10  & $2.4 \ \times 10^{-12}$ \\ \hline
\end{tabular} \\
\end{tabular}
\caption{Limitation of SGWB cosmological amplitude measurement, modification of the acceleration noise and the optical noise. $\Omega_{0,lim}$ is given for error level at $20 \%$ ($\frac{\Delta \Omega_0}{\Omega_0} = 0.2$). This is the upper horizontal dashed line in  Figs~\ref{fig:Uncertaintyacc}~and~\ref{fig:Uncertaintypos}.}
\label{tab:Uncertaintyaccpos}
\end{table}

\section{Conclusion} 
\label{sec:conclusion}

The study presented in this paper addresses the effects of modifications to the LISA noise, and their impact on spectral separation of SGWB signals. The initial part of the study consisted of estimating the various noise sources that were correlated to the differential acceleration of the test mass measured in the LPF mission. It is reduced to the study of the variations of amplitudes common to the differential acceleration of the tests-mass of LPF with three other channels. Respectively, with the LPF data we observed the effects of orbital compensation with the $\mu$ thrusters, the effects of variations of temperature, and magnetic field noise. Significant correlations in LPF data have been observed. 
The projection of these LPF correlations to the LISA mission corresponds to a worst-case scenario.

We have then introduced the correlations observed in LPF into the LISA TDI algorithm. Studies using the Fisher Information and a Bayesian MCMC method establish a limiting estimate for the LISA observation of the SGWB of cosmological origin. We set a limit of the detectability with our spectral separation algorithm at 20\% error for the cosmological SGWB measurement. 
The cosmological SGWB detection performance is degraded by 22\%. This modification of the LISA noise and subsequent observation of the effects on measuring the LISA cosmological SGWB can be seen as a type of figure of merit study.  

In order to understand the effects of an increase in the LISA noise compared to the LISA noise model, the last section of the paper presents the examination of different cases of LISA noise and their consequences for the LISA SGWB observation. Many noise sources could increase the initial noise budget of LISA, as the correlated noise studied with LPF has demonstrated. 
Simulations and studies of LISA noise are certainly necessary for estimating the scientific possibilities for the LISA mission. The examination of LPF data, as demonstrated here and in a study with a similar philosophy~\cite{PhysRevD.105.042002}, offers a means to estimate possible deleterious scenarios that LISA might encounter.

\section*{Acknowledgements}
\label{sec:awk}
The G. B. and N. C. thank the Centre national d'études spatiales (CNES) for support for this research. 
N. C. and R. M. acknowledge support by  Marsden Fund Grant No. 21-UOA-105 from Government funding, administered by the Royal Society Te Ap\={a}rangi.
\section*{Data Availability}

The data generated for our study presented in this article will be shared upon reasonable request to the corresponding author.
\nocite{*}

\bibliography{bibli}

\end{document}